\documentclass{article}
\usepackage[english]{babel}
\usepackage[letterpaper,top=2cm,bottom=2cm,left=3cm,right=3cm,marginparwidth=1.75cm]{geometry}

\usepackage{amsmath}
\usepackage{graphicx}
\usepackage[colorlinks=true, allcolors=blue]{hyperref}
\usepackage{authblk}
\usepackage[numbers, sort&compress]{natbib}
\usepackage{caption}
\usepackage{subcaption}
\usepackage{doi}

\title{Unequal changes in commuting patterns across socio-economic strata in response to pandemic restrictions}
\author[1]{Cristiano Marinelli}
\affil[1]{ISI Foundation, Turin, Italy}
\author[1,2]{Leo Ferres}
\affil[2]{Data Science Institute, Universidad del Desarrollo, Santiago, Chile}
\author[3]{Niccolò Comini}
\affil[3]{The World Bank, Washington, DC, USA}
\author[1,3]{Nicolò Gozzi}
\author[4,3,5]{Nicola Perra}
\affil[4]{School of Mathematical Sciences, Queen Mary University of London, London, UK}
\affil[5]{Alan Turing Institute, London, UK}

\begin{document}
\maketitle

\begin{abstract}
Commuting patterns are a central component of urban dynamics and many societal activities. Exogenous shocks, such as a pandemic, might drastically modify them inducing heterogeneous variations across socioeconomic strata. Here, we quantify changes in work commuting patterns in Bogotá, Colombia during three different periods of the COVID-19 pandemic: pre-pandemic ($2019$), COVID-19 restrictions ($2020$), and partial reopening ($2021$). To this end, we use anonymized mobile phone data to infer home and work locations from recurring nighttime and weekday connection patterns, and to build daily commuting metrics. We aggregate mobility flows by administrative boundaries and socioeconomic strata. Additionally, we enrich the dataset with a range of other variables such as territorial vocation (i.e., urban versus rural), demographic information (i.e., population density) and, as a proxy for digital infrastructure quality, geolocated Speedtest\textsuperscript{\tiny\textregistered} measurements from Ookla\textsuperscript{\tiny\textregistered}. We find a marked reduction of commuting during restrictions in $2020$ and a strong recovery in $2021$, but with persistent heterogeneity across socioeconomic strata. Indeed, while commuting declined similarly across income groups during restrictions, groups of the population in the lower-income bracket rebounded faster to pre-pandemic levels. On the contrary, we find that groups in the higher-income bracket managed to keep higher stay-at-home behavior. Regression analyses reveal that territorial characteristics and disparities in digital connectivity significantly contribute to these differences, suggesting that infrastructure investments could help mitigate mobility-based inequalities. 
\end{abstract}

\section{Introduction}

Human mobility is a fundamental driver of social, economic, and environmental processes across spatio-temporal scales. Movements are critical for social interactions, access to job opportunities, urbanization, and the spread of infectious diseases~\cite{MobilityUrbanTheory2025,HumanMobilityDRR_IOM,pereyra2019mobility,peixoto2020role,balcan2009multiscale}. Mobility patterns are complex and dynamic, yet they exhibit repeating structures that shape society, such as commuting to work~\cite{gonzalez2008understanding,song2010c,christianm.schneider2013,kung2014,pappalardo2015,alessandretti2020}.  

At the same time, the way we move is influenced by exogenous factors that interact heterogeneously  with local conditions across socio-economic strata. Large-scale emergencies, like pandemics, might abruptly disrupt mobility flows, creating natural experiments to investigate the interplay between such external shocks and local contexts in different groups of the population~\cite{perra2021non,flaxman2020estimating,snoeijer2021measuring,haug2020ranking,cowling2020impact,bonaccorsi2020economic,skarp2021systematic,klein2020reshaping,gozzi2021estimating}. Indeed, during the COVID-19 Pandemic, the drastic contraction in mobility we witnessed had far-reaching and unequal socioeconomic consequences. Restrictions on movement and commercial activity triggered one of the largest employment shocks in recent history, resulting in income losses and reduced purchasing power that reinforced existing and structural inequities~\cite{saadi2020vicious,brodeur2021literature,permal2022cascading,robinson2021cascading}. Decline in public transport usage was particularly pronounced, as it was directly targeted by containment policies~\cite{bucsky2020modal,Arellana2020covid,google2020covid,aloi2020effects}. Since public transport users are disproportionately drawn from lower-income groups, these groups were among the most severely affected by mobility restrictions. At the same time, essential workers—who continued commuting to/from work to sustain critical economic functions—were also predominantly drawn from lower-income social strata. Evidence suggests that mobility reductions were less pronounced among these groups~\cite{bargain2021poverty,yabe2020non,30314,gozzi2021estimating}. Despite these unequal impacts, the temporary reduction in commuting also produced short-lived positive externalities, including sharp declines in emissions and relief for workers previously exposed to long commuting times, which are known to be associated with lower well-being and poorer mental health outcomes~\cite{nikolaeva2023living,clark2020commuting,lorenz2018does}. More broadly, the disruption of commuting has fueled discussions on proximity-based urban planning approaches aimed at reducing long daily commutes and improving urban well-being~\cite{Allam2020}.

These dynamics are particularly salient in cities of the Global South, where high rates of informal and often precarious employment are compounded by rapid urbanization and high population densities. In Latin America, approximately $80\%$ of the population resides in urban areas, making cities the primary arena in which socioeconomic inequalities and mobility constraints intersect~\cite{Roberts2017}. The urban geographies of the region are characterized by pronounced spatial segregation driven by differences in income, socioeconomic stratification, and ethnicity, which correlate with unequal access to housing, infrastructure, and local amenities. Nearly a quarter of the urban population lives in dense informal settlements~\cite{Inostroza2017}. At the same time, the concentration of economic activities and high land values in a limited number of central areas, combined with sustained demographic growth, have contributed to the expansion of low-income populations in peripheral zones. As a result, the urban structure of many Latin American cities generates disproportionately long travel times for populations that rely primarily on public transport and active modes, which together account for a large share of daily trips~\cite{EstupinanRodriguez2008}. During the COVID-19 Pandemic, these structural conditions amplified the distributional effects of mobility restrictions, particularly in contexts where reductions in public transport supply increased waiting times, congestion, and social tensions in high-density urban areas~\cite{Arellana2020covid}.

Understanding how commuting patterns respond to large-scale disruptions has traditionally relied on travel surveys and census data. While surveys provide rich socioeconomic detail, they are costly, infrequent, and subject to sampling and cognitive biases. In recent years, the increasing availability of mobile phone data, such as Call Detail Records (CDR) and eXtended Data Records (XDRs), has enabled researchers to observe population mobility with unprecedented spatial and temporal resolution. Despite ongoing methodological challenges related to the inference of trips and interactions from sparse network connections, these data have been widely used to study commuting flows and behavioral responses to policy interventions ~\cite{kouam2025beyond,klein2020reshaping}.

Urban mobility research has long examined how movement patterns relate to the physical structure of cities, transportation networks, population density, and socioeconomic characteristics. More recently, attention has also turned to the role of digital infrastructure, particularly Internet accessibility, as a factor shaping mobility behavior and the feasibility of remote work~\cite{gozzi2023adoption,gozzi2024bridging,oecd2021bridging,SAXON2022101874}. However, empirical evidence based on high-resolution mobility data remains heavily concentrated in developed countries, leaving important gaps in our understanding of commuting dynamics in developing regions.

In this context, here we focus on commuting patterns—defined as regular movements between inferred home and work locations—in the metropolitan area of Bogotá, Colombia between $2019$ and $2021$. Bogotá represents a particularly relevant case study: it is a large and highly urbanized Latin American city that has implemented substantial mobility and social policies, yet continues to exhibit marked socioeconomic inequalities~\cite{guzman2021covid}. The city also features a well-established socioeconomic stratification system at the cadastral-block level, which provides a reliable proxy for residents’ income 
~\cite{guzman2019strata}.

We analyze anonymized mobile phone XDR data covering three distinct periods: a pre-pandemic baseline, the phase of strict COVID-19 mobility restrictions in $2020$, and the $2021$ period characterized by partial reopening. To enrich the mobility information, we integrate these data with multiple external sources capturing socioeconomic stratification, territorial vocation (i.e., urban versus rural), demographic information, and measures of digital infrastructure quality derived from Internet Speedtest measurements provided by Ookla. Overall, the data allow us to examine how commuting patterns evolved across different socioeconomic groups, spatial contexts, and stages of the pandemic.

Our contribution is threefold. First, we provide new empirical evidence on commuting dynamics during crises in a major Latin American city using high-resolution mobile phone data. Second, we assess how socioeconomic inequality, spatial segregation, and urban structure shape heterogeneous mobility responses to large-scale disruptions across the population. Third, we highlight the role of digital infrastructure as a complementary dimension influencing commuting behavior in contexts marked by strong social and spatial heterogeneity. This study contributes to a more detailed understanding of urban mobility and offers insights relevant for evidence-based transport and social policy in developing urban regions.

\section{Results}
\subsection{Changes in commuting patterns during pandemic restrictions}

We analyze anonymised eXtended Data Records (XDRs) obtained from a major Mobile Network Operator (MNO) in Bogotá, Colombia. The data covers three months (March, April, and May) in each of three years ($2019$, $2020$, and $2021$). XDRs capture network-level connection events generated whenever a mobile device interacts with the cellular infrastructure, such as during data usage, app activity, or periodic signaling. We refer the reader to the Materials and Methods section (section~\ref{sec:mem}) for more details about the data.

Using this data we study how commuting behavior evolved in Bogotá over our study period. To this end, we adopt a mobility metric designed to capture changes in estimated daily home-to-work connections. This metric, which we call \textit{non-commuting fraction} ($p_{h,w,y}$), is defined for each home–work location pair and it is computed as the fraction of estimated commuters during working hours. Intuitively, $p_{h,w,y}(t)=0$ indicates that all users living in $h$ and working in $w$ commuted to work on day $t$ of year period $y$, while $p_{h,w,y}(t)=1$ indicates that they all stayed home. Full details on metric calculation and are provided in section~\ref{sec:mem}. 

\begin{figure}[ht]
    \centering
    \includegraphics[width=1\textwidth]{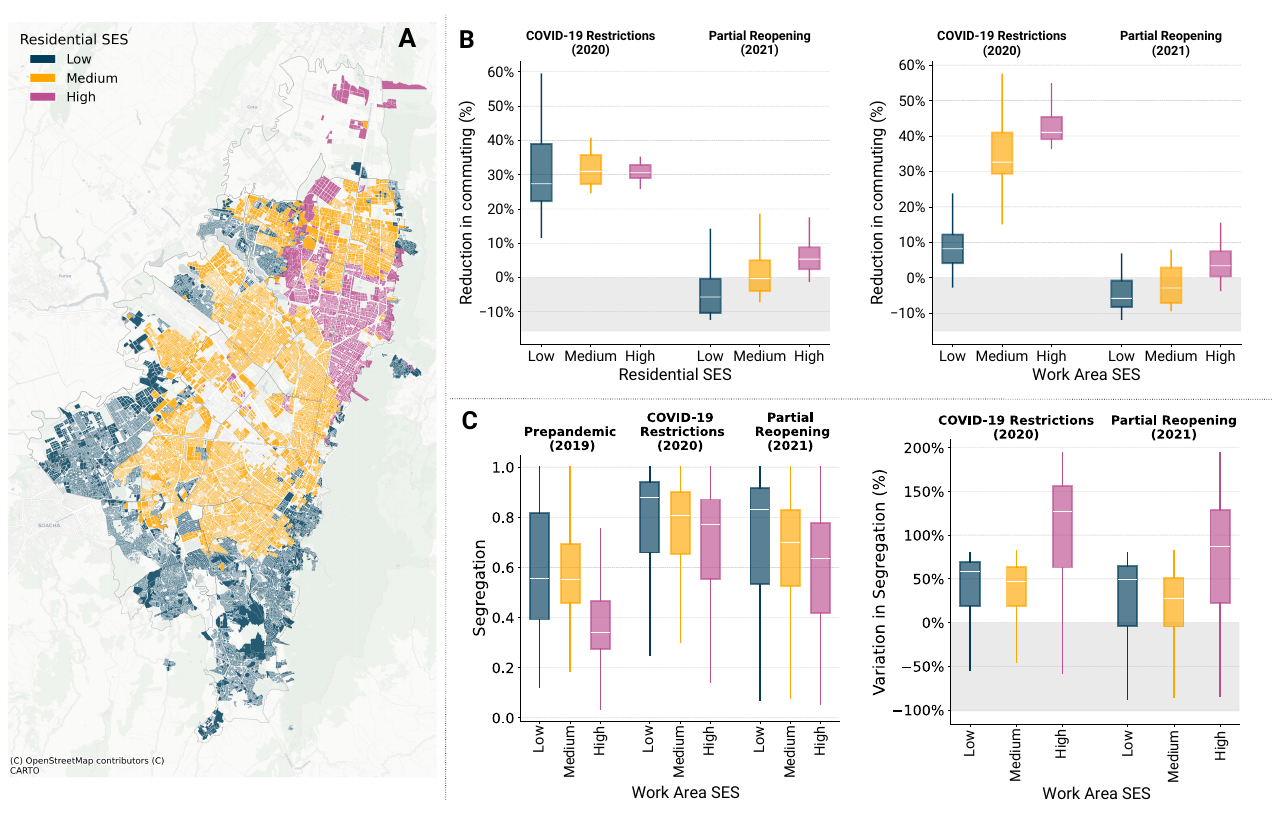}
    \caption{\textbf{Spatial distribution of socioeconomic stratification and changes in commuting patterns across residential and workplace areas}. \textbf{(A)} Map of Bogotá showing residential blocks colored by socioeconomic stratum (SES) (low in blue, medium in yellow, and high in purple). Black lines indicate local planning unit (UPL) boundaries. \textbf{(B)} Reduction in commuting relative to the $2019$ baseline, stratified by residential SES (left) and work area SES (right) during COVID-19 restrictions ($2020$) and partial reopening ($2021$). \textbf{(C)}. Workplace segregation patterns across periods. Left: Distribution of segregation indices by workplace SES for pre-pandemic ($2019$), COVID-19 restrictions ($2020$), and partial reopening ($2021$) periods. The segregation index ranges from $0$ (i.e., perfect socioeconomic mixing) to $1$ (i.e., complete segregation). Right: Variation in segregation relative to the $2019$ baseline, showing percent change in segregation by workplace SES. In all plots, the box boundaries represent the interquartile range ($IQR$) between the first and third quartiles ($Q1$ and $Q3$), the line inside the box indicates the median, and the whiskers extend to $1.5$ times the $IQR$ from the quartiles.}
    \label{fig:panel1}
\end{figure}

Our spatial units of analysis are defined by subdividing Bogotá's local planning units (UPLs) according to socioeconomic stratification (SES). Each residential block within a UPL is assigned an SES level (low, medium, or high). We aggregate all blocks within each UPL that share the same SES classification, creating sub-UPL zones characterized by SES homogeneity. This process yields $56$ geographic regions across the city, each representing a specific UPL-SES combination. Fig.~\ref{fig:panel1}A shows the spatial distribution of socioeconomic stratification across residential blocks (colored by SES level) and UPL (black boundaries) in Bogotá. The spatial pattern reveals pronounced North-South segregation, with low-SES populations concentrated in Southern and peripheral areas and high-SES populations in Northern sectors. Further details on the stratification system and spatial aggregation are provided in section~\ref{sec:mem}.

Fig.~\ref{fig:panel1}B shows changes in commuting rates during the strict restriction period ($2020$) and partial reopening ($2021$) relative to the pre-pandemic baseline, stratified by SES of home locations (left panel) and work locations (right panel). When considering home location SES (see Fig.~\ref{fig:panel1}B, left panel), we find that during $2020$ restrictions, commuting declined by approximately $30\%$ across all residential SES groups. By $2021$, commuting largely recovered to pre-pandemic levels, but with significant heterogeneity across income groups: residents experiencing lower-SES returned to baseline commuting rates, while residents experiencing higher-SES sustained higher work-from-home behavior (difference statistically significant by median permutation and Mann-Whitney U test). 

Unlike residential patterns, workplace SES reveals sharp gradients throughout both periods (see Fig.~\ref{fig:panel1}B, right panel). During $2020$ restrictions, high-SES work areas experienced approximately $40\%$ reductions in worker presence, compared to only $10\%$ reductions in low-SES work areas. This gradient persisted into $2021$, though attenuated. Combined, these results indicate that the nature of employment locations and not just individual characteristics fundamentally shapes commuting variations, with jobs in high-SES areas offering greater flexibility for remote work.

This asymmetric reduction across workplace SES levels is also reflected in spatial segregation dynamics. The left panel of Fig.~\ref{fig:panel1}C shows the distribution of workplace segregation indices across different SES levels. The segregation index ranges from $0$ (perfect mixing, where a workplace draws workers equally from all residential SES groups) to $1$ (complete segregation, where workers come exclusively from a single residential SES). Across all periods, high-SES workplaces exhibit lower baseline segregation, while low-SES workplaces show the highest levels. However, the right panel of Fig.~\ref{fig:panel1}C reveals that segregation increased across all workplace SES categories in both $2020$ and $2021$ relative to pre-pandemic baseline, with the largest increases occurring in high-SES workplaces. In other words, while high-SES employment areas remained less segregated in absolute terms, they experienced the sharpest rise in socioeconomic homogeneity during and after partial relaxation of pandemic restrictions.

Beyond changes in commuting frequency, we also observe significant socioeconomic disparities in commuting distance. In the Supplementary Information we show the distribution of commuting lengths (in kilometers) across SES groups. While average commuting distances declined from baseline levels across all groups, the magnitude of reduction exhibits a strong socioeconomic gradient: individuals experiencing higher-SES substantially reduced their commuting distances, whereas workers experiencing lower-SES showed smaller changes in trip length. This pattern suggests that lower-income workers not only returned to in-person work more quickly but also maintained longer commutes, potentially reflecting reduced flexibility in both workplace location and transportation options.

\subsection{Access to and quality of the digital infrastructure}

As mentioned, we include Internet Speedtest data from Ookla in our analysis. The data is used as proxy of the digital infrastructure landscape across socioeconomic groups~\cite{gozzi2023adoption,gozzi2024bridging}. We investigate two measures of digital infrastructure quality. First, we study the median fixed broadband download speed in the home area as a proxy for residential Internet quality, hypothesizing that better home connectivity facilitates remote work adoption~\cite{gozzi2024bridging}. Second, we investigate the log-ratio of home-to-workplace Internet speeds, capturing relative connectivity conditions. This measures whether individuals experience substantially better connectivity at work versus home. Here, we hypothesize that larger disparities in favor of the workplace incentivize in-person commuting, while similar or superior home connectivity supports and facilitates remote work.

Fig.~\ref{fig:panel2}A shows the distribution of home download speeds across residential SES levels, revealing a pronounced digital divide: high-SES areas exhibit median speeds higher than $50$ Mbps, while low-SES areas of around $10-20$ Mbps. Medium-SES areas show intermediate values. Fig.~\ref{fig:panel2}B shows the relationship between absolute home download speed and the log speed differential (i.e., home minus work) across all home-work pairs, colored by residential SES. The plot reveals that higher absolute home speeds are generally associated with positive speed differentials (better connectivity at home than work), while low-SES individuals (blue) tend to cluster in regions with lower absolute speeds and negative speed differentials, suggesting compounding disadvantages in both home infrastructure quality and relative workplace connectivity. Despite these clear median trends, we acknowledge substantial within-group variation exists in both absolute and relative Internet speed, supporting the inclusion of both variables in the following analysis.

\subsection{Predictors of commuting behavior changes}

To identify the predictors of the heterogeneous changes in commuting observed, we develop a multivariate regression model. For each home-work pair and period (i.e., baseline, restriction, partial reopening), we compute the median non-commuting fraction across all available working days. The dependent variable is the change in this fraction relative to the pre-pandemic baseline for the same home-work pair.
The model includes a range of covariates characterizing the SES level in residential and workplace areas, territorial vocation (i.e., urban vs. non-urban), resident population, gender ratio, and commuting distance. We additionally incorporate the two measures of digital infrastructure quality derived from Ookla Speedtest data described in the previous section. All numeric variables are standardized to zero mean and unit variance. Categorical variables are encoded using one-hot encoding, with one reference category. In Tab.~\ref{tab:covariates} we provide full list of covariates with their interpretation. More details on their definition and sources are provided in Section~\ref{sec:mem}. Additionally, in the Supplementary Information, we estimate simple and partial correlations among covariates and the dependent variable as well as univariate regression models. Results obtained in that context support the definition of the multivariate regression model and the findings presented below. 

\begin{table}[htpb]
\centering
\begin{tabular}{llp{6cm}l}
\hline
\textbf{Variable} & \textbf{Type} & \textbf{Description} & \textbf{Source} \\ \hline
\multicolumn{4}{l}{\textbf{Dependent Variable}} \\
$\Delta$ Non-commuting & Continuous & Change in median non-commuting fraction relative to 2019 baseline & XDR data \\ 
\hline
\multicolumn{4}{l}{\textbf{Independent Variables}} \\
Period (2021) & Categorical & Dummy variable for partial reopening (ref: 2020 restrictions) & / \\
Residential SES (Low) & Categorical & Dummy variable for low home area SES (ref: High) & Government data \footnotemark\\
Residential SES (Medium) & Categorical & Dummy variable for medium home area SES (ref: High) & Government data  \footnotemark[\value{footnote}]\\
Work Area SES (Low) & Categorical & Dummy variable for low work area SES (ref: High) & Government data \footnotemark[\value{footnote}] \\
Work Area SES (Medium) & Categorical & Dummy variable for medium work area SES (ref: High) & Government data \footnotemark[\value{footnote}] \\
Home Vocation (Urban) & Categorical & Dummy variable for urban home area (ref: non-urban) & Government data \footnotemark[\value{footnote}] \\
Work Vocation (Urban) & Categorical & Dummy variable for urban workplace (ref: non-urban) & Government data \footnotemark[\value{footnote}] \\
Commuting Distance & Continuous & Median distance (km) between home and work areas & Computed \\
Population & Continuous & Total residential population in home area & XDR data \\
Gender Ratio & Continuous & Female-to-Male ratio in home area & XDR data \\
Home Download Speed & Continuous & Median broadband download speed (Mbps) in home area & Ookla \\
Log Speed Ratio & Continuous & Log(home speed / work speed) & Ookla \\ \hline
\end{tabular}
\caption{Variables used in the multivariate regression.}
\label{tab:covariates}
\end{table}
\footnotetext{Bogot\'a open data portal: \url{https://datosabiertos.bogota.gov.co/dataset/manzana}}

\begin{figure}[ht]
    \centering
    \includegraphics[width=1\textwidth]{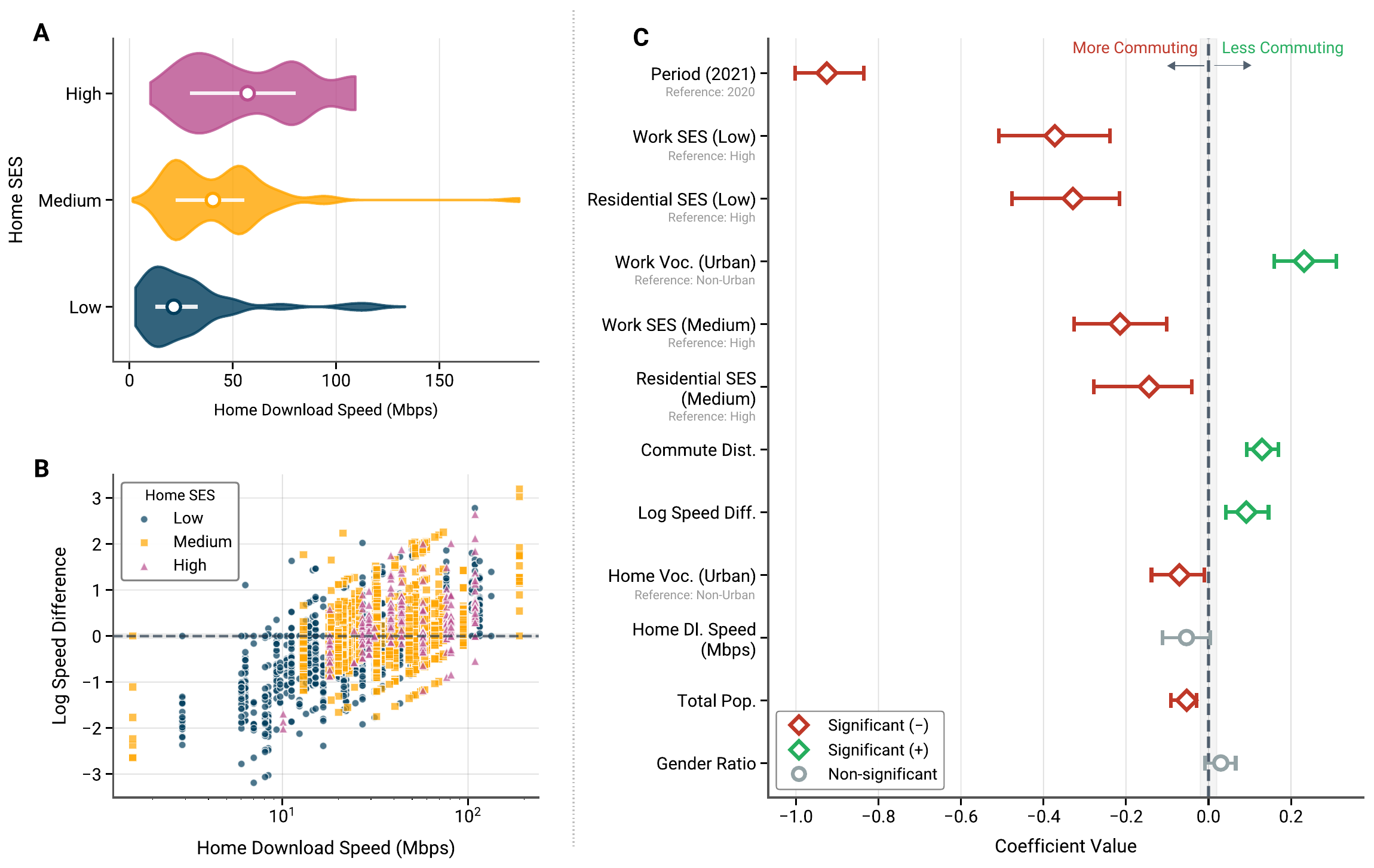}

    \caption{\textbf{Internet connectivity patterns and regression coefficients}. \textbf{(A)} Home download speed distributions across residential SES groups. \textbf{(B)} Relationship between home download speed and speed differential (home-work) by SES level. Different markers and colors represent SES groups. \textbf{(C)} Regression coefficients with $95\%$ confidence intervals, ranked by absolute magnitude. Red/green indicates significant effects (negative and positive) at the $5\%$ confidence level; gray indicates non-significant effects.}
    \label{fig:panel2}
\end{figure}

We estimate a multivariate linear regression model using ordinary least squares (OLS) with bootstrapped confidence intervals. Figure~\ref{fig:panel2}C presents median regression coefficients with $95\%$ confidence intervals for all covariates. Positive coefficients indicate increased non-commuting fraction (i.e., reduced commuting), while negative coefficients indicate increased commuting. The period dummy variable shows the largest effect. Using $2020$ restriction period as reference, the $2021$ partial reopening exhibits significantly higher commuting propensity, consistent with descriptive findings of previous section. Both residential and workplace SES significantly shape commuting patterns. Relative to high SES reference areas, low and medium residential SES show negative coefficients (low SES more strongly negative), indicating an association with smaller reductions in commuting. Workplace SES exhibits similar patterns: low and medium SES work areas are associated with smaller commuting reductions than high-SES employment zones. Urban workplaces are associated greater commuting reductions than non-urban locations (positive, significant coefficient), while commuting distance shows a positive association, with longer-distance commuters reducing travel more after controlling for other factors. On the contrary, urban residential areas are associated with increased commuting with respect to non-urban home areas. Population exhibits a negative relationship, indicating that less populated areas are associated with larger commuting reductions. Regarding digital infrastructure, absolute home Internet speed shows no significant effect. However, the log-ratio of home-to-workplace speeds is positive and significant: better home connectivity relative to workplace is associated with reduced commuting, while superior workplace connectivity corresponds to increased commuting. Finally, gender ratio shows no significant relationship with commuting changes. The multivariate regression achieves a median $R^2$ of $0.303$, with a $95\%$ bootstrap confidence interval of $[0.272,\,0.336]$. In the Supplementary Information, we repeat the analysis using random forests and AdaBoost regression methods. Both lead to results consistent with those presented here.

\section{Conclusions}

The COVID-19 Pandemic, together with the widespread adoption of mobile technologies, has created unprecedented opportunities to study changes to human mobility during crises, at scale. In this work, we used anonymized mobile phone data to investigate how commuting-to-work behavior changed during the Pandemic in Bogotá, one of the largest metropolitan areas in South America. The temporal coverage of the data allowed us to compare pre-pandemic conditions in $2019$, the period of strict mobility restrictions in $2020$, and the subsequent phase of partial reopening in $2021$.

Our results show a sharp and abrupt reduction in commuting activity during the restriction period of $2020$, followed by a substantial recovery in $2021$, when commuting flows relaxed towards pre-pandemic levels. However, this recovery was not uniform across the population. 
In particular, individuals associated with lower socioeconomic strata exhibited a smaller reduction in commuting during the restriction period and, in the later phase, a sharper increase in commuting relative to higher-income groups. In contrast, higher-income populations displayed a greater capacity to reduce commuting even outside of the strict phase of the restrictions, especially toward highly urbanized and central workplace areas. These patterns are consistent with differences in occupational structure and access to remote or hybrid working arrangements, which are typically more prevalent among higher-income workers and in office-based activities concentrated in central urban areas. Commuting distance also emerged as another factor shaping mobility responses. In a spatially extended city such as Bogotá, characterized by intense traffic congestion, particularly during peak hours, longer commuting distances were associated with larger reductions in commuting activity. Finally, digital infrastructure played a significant role: although absolute internet speed in residential area was not a significant predictor of changes in commuting behavior, a higher relative difference in connectivity between home and workplace locations was associated with stronger reduction in mobility, highlighting the importance of access to reliable digital infrastructure in shaping commuting behavior.

The present work comes with limitations. First, the mobility dataset employed in this analysis captures only individuals with mobile phones and active connections to the network, which may skew toward certain demographic groups. Nonetheless, the data provider's significant market share in Colombia partly mitigates this concern. Home and work locations were inferred from mobility patterns using standard location detection algorithms, as detailed in section~\ref{sec:mem}, with location assignments rerun annually to capture changes in residence and workplace across years. Nonetheless, we acknowledge that the algorithm may fail to detect rapid transitions. For instance, individuals who lost employment within one of the three study period would continue to be classified as workers, with their subsequent stay-at-home behavior attributed to remote work rather than unemployment. More broadly, our measure of commuting reduction is a proxy: we cannot definitively distinguish whether individuals staying at home were working remotely, on leave, or unemployed during a given period.
Second, the data used to estimate the quality of and access to the digital infrastructure represents user-initiated speed tests and may not fully capture actual connectivity experienced by all residents. The measurements reflect voluntary testing behavior, which may correlate with technical proficiency and could introduce selection bias in infrastructure quality estimates~\cite{feamster2020measuring}. Nevertheless, Ookla Speedtest data is widely adopted by academic and governmental institutions as a standard measure of internet connectivity~\cite{feamster2020measuring,ford2021form,gozzi2023adoption,gozzi2024bridging}. Moreover, our analytical approach mitigates potential biases from heterogeneous testing behavior, as detailed in section~\ref{sec:mem} and refs.~\cite{gozzi2024bridging, gozzi2023adoption}.
Third, our regression analysis identifies associations rather than causal relationships. While we control for multiple covariates, unobserved factors such as industry sector composition, employer policies, household composition, or childcare responsibilities, may partially confound some of the observed relationships between SES, digital infrastructure, and commuting behavior.
Finally, our findings are derived in the context of Bogotá during the COVID-19 pandemic and may not generalize to other urban contexts with different spatial structures, policy responses, labor market compositions, or digital infrastructure landscapes.

In conclusion, our work contributes to the growing literature devoted at understanding how pandemic-induced disruptions may amplify pre-existing urban inequalities through the lens of mobility and digital infrastructure. Our results confirm the key role of socio-economic disparities in shaping the mobility response of different populations during crises. They reinforce the need for policies aimed at addressing such inequalities, for example via targeted interventions supporting individuals that have lower capabilities to reduce commuting, such as those living in lower socio-economic conditions or in specific vocational areas. Additionally, our findings highlight how fast internet connectivity might be crucial to enhance societal resilience during major exogenous shocks such as a health emergency and call for policies devoted to support equal digital access across socio-economic strata. Furthermore, our work showcases the value of Data for Good initiatives of private corporations to help probe, monitor, and study different aspects of human behavior at scale, while upholding the highest standards of privacy.

\section{Materials and methods}
\label{sec:mem}

\subsection{XDR data}
We analyzed mobile phone connection data from a major Mobile Network Operator (MNO) covering three months (March, April, and May) in each of three years ($2019$, $2020$, and $2021$) in Bogotá, Colombia. The MNO had approximately $25\%$ of the national market share at the time of data collection. The dataset consists of anonymized eXtended Data Records (XDRs), which capture network-level connection events generated whenever a mobile device interacts with the cellular infrastructure, such as during data usage, app activity, or periodic signaling. Compared to traditional Call Detail Records, XDRs provide higher temporal resolution and a more comprehensive representation of user presence and activity, making them particularly well suited for the study of routine mobility patterns and commuting behavior \cite{gonzalez2008understanding,berrios2024near,gozzi2021estimating,josiane2025characterizing}.

The dataset contained $7,686,507,384$ connection events across the three-year period. In $2019$, we analyzed $2,813,986,462$ connections from $253,653$ subscribers ($140,168$ male, $113,485$ female). In $2020$, we analyzed $1,965,040,940$ connections from $224,180$ subscribers ($125,287$ male, $98,893$ female). In $2021$, we analyzed $2,907,479,982$ connections from $275,782$ subscribers ($155,392$ male, $120,390$ female). Each connection event recorded the user identifier (anonymized), cell tower identifier, tower coordinates (latitude and longitude), connection timestamp, and user gender. It is important to mention and stress how the analysis reported here are based entirely on spatially aggregated data rather than individual users' trajectories. 

\subsection{Commuting patterns: home and work detection}

We focus on commuting patterns defined as regular movements between home and work locations. Home locations were inferred from nighttime tower connections between $22{:}00$ and $06{:}00$. For each user, we identified the cellular tower with the highest number of nighttime connections across the three-month observation period within each year, following established methodologies~\cite{pappalardo2021evaluation}. Work locations were inferred from weekday tower connections during business hours (Monday through Friday). A work tower was assigned to a user if they connected to that tower on at least three different weekdays within the selected work-hour window. users who did not meet these criteria for either home or work identification were excluded from the analysis.

To account for heterogeneity in work schedules across the population, we considered three alternative definitions of working hours: $07{:}00-17{:}00$, $08{:}00-17{:}00$, and $09{:}00-17{:}00$. For each definition, home--work locations and commuting patterns were inferred independently. Unless otherwise stated, the analysis presented in the main text the results refer to the $09{:}00-17{:}00$ time window; results obtained using the other definitions are qualitatively similar and are reported in the Supplementary Materials.

A commuting event is defined as a day in which a user is observed connecting to both their inferred home tower $h$ and work tower $w$. The resulting sets of home and work towers have cardinalities $|H|=1432$ and $|W|=1441$, respectively, with only $12$ towers belonging exclusively to one set. For each date $t$, home--work tower pair $(h,w)$, and year $y$, we compute four basic quantities: the number of male users who commuted $m_{h,w,y}(t)$, the total number of male users associated with the pair $m_{h,w,y}$, the number of female users who commuted $f_{h,w,y}(t)$, and the total number of female users associated with the pair $f_{h,w,y}$. From these, we derive the total number of commuters at time t:
\[
n_{h,w,y}(t) = m_{h,w,y}(t) + f_{h,w,y}(t),
\]
and the total population associated with each home--work pair,
\[
n_{h,w,y} = m_{h,w,y} + f_{h,w,y}.
\]
These latter quantities are constant within each year but may differ across years. To reduce spurious detections, we exclude pairs for which $h=w$, as commuting cannot be reliably identified in these cases.

After preprocessing, the final dataset contains $79,031$ unique home--work tower pairs and $7,270,852$ daily observations in $2019$, $31,676$ pairs and $2,914,192$ daily observations in $2020$, and $51,250$ pairs and $4,715,000$ daily observations in $2021$, spanning a total of $276$ days across the three years.

From these quantities, we define the \textit{non-commuting fraction}
\begin{equation}
p_{h,w,y}(t) = 1 - \frac{n_{h,w,y}(t)}{n_{h,w,y}},
\end{equation}
which measures the fraction of individuals associated with a given home--work pair who did not commute on day $t$ in year $y$. This metric ranges between zero and one and increases with the share of the population remaining at home during working hours. In particular, $p_{h,w}(t)=1$ indicates that none of the individuals usually commuting between $h$ and $w$ were observed traveling on day $t$.

To facilitate comparisons across time and reduce the impact of systematic labeling noise in home and work inference, we also consider a relative variation with respect to a pre-pandemic baseline. Specifically, we define
\begin{equation}
r_{h,w,y}(t) = \frac{p_{h,w,y}(t)}{\mathrm{med}\left[p_{h,w,y}(t \in B)\right]},
\end{equation}
where $\mathrm{med}[\cdot]$ denotes the median and $B$ identifies the baseline period. We partition the data into three phases: (i) a pre-pandemic baseline period, comprising all of $2019$ and the portion of $2020$ preceding March $17$th ($71$ days after data cleaning); (ii) a restriction period from March $17$th to May $31$th, $2020$, corresponding to the strictest mobility containment measures ($43$ days); and (iii) a post-restriction period covering $2021$ ($58$ days). The baseline period captures habitual commuting behavior and serves as the reference against which changes during and after restrictions are evaluated.

Finally, the definition of $p_{h,w}(t)$ naturally allows for aggregations beyond the tower level. While cellular towers represent the finest spatial resolution of the dataset, they can be grouped into larger spatial units. Let $A_h$ and $A_w$ denote two generic partitions of the urban space into spatial areas used to aggregate home and work locations, respectively. These partitions are not unique, as different analyses may rely on different spatial classifications (e.g., socioeconomic strata, administrative units, territorial vocation, or their combinations). Each element $a_h \in A_h$ and $a_w \in A_w$ corresponds to a spatial area defined by common characteristics and contains the set of cellular towers located within its boundary, which are aggregated in the analysis.
Given these sets, we define aggregated commuting flows as:
\begin{align}
n_{a_h,a_w}(t) &= \sum_{h \in a_h,\, w \in a_w} n_{h,w,y}(t),\\
n_{a_h,a_w,y} &= \sum_{h \in a_h,\, w \in a_w} n_{h,w,y},
\end{align}
where, in contrast to the tower-level case, it is possible that $a_h = a_w$. The aggregation strategies and associated datasets are introduced in the next subsection.

\subsection{Other data}

As mentioned, our analyses are conducted considering a range of other datasets.\\

\textbf{Socioeconomic stratification data}. In Colombia, residential urban blocks (i.e., \textit{manzanas}) are officially classified by the government into six socioeconomic strata, ranging from 1 (lowest) to 6 (highest) \cite{Heroy2021}. While this classification is not uniformly representative of income across the country, previous studies based on surveys and interviews have shown that, in the specific case of Bogotá, socioeconomic strata are strongly correlated with household income and broader socioeconomic status~\cite{guzman2019strata}. In particular, strata 1 and 2 correspond to low-income households, strata 3 and 4 to middle-income households, and strata 5 and 6 to high-income households. This sharp stratification is a distinctive feature of Bogotá.

The Bogotá municipal government makes these data publicly available. We use it as a proxy to classify the commuting population into three income groups (i.e., low, medium, and high). Using the spatial correspondence between cellular towers and nearby residential blocks, we assign to each tower the socioeconomic strata of the blocks it predominantly serves. Through this procedure, each commuter is associated with both a home and a work income level. The home income level is interpreted as a proxy for the individual socioeconomic status of commuters, whereas the work income level reflects the socioeconomic characteristics of the destination area.

After this assignment, the commuting population is composed of approximately $28\%$ low-income, $55\%$ middle-income, and $17\%$ high-income individuals. Income level thus becomes a natural aggregation dimension for the analysis.

Before producing the income-stratified data, we restrict the analysis to working days by excluding weekends and public holidays. To facilitate comparisons across different phases of the pandemic, the data are further partitioned into three distinct periods, as described above.

The analyses shown in Fig.~\ref{fig:panel1}B are obtained by aggregating home–work commuting flows over areas defined by socioeconomic strata. For the residential-based representation, commuting measures are aggregated by grouping flows according to the socioeconomic status of the home area, summing over all associated workplace destinations. Conversely, the workplace-based representation aggregates flows according to the socioeconomic status of the work area, summing over all residential origins. These two complementary aggregations allow us to distinguish how commuting reductions vary with respect to the socioeconomic characteristics of residential locations versus employment locations.\\

\textbf{Commuting distance.} Distance to work represents another important dimension for characterizing mobility behavior. While commuting patterns are influenced by infrastructure, public transport availability, and urban services, the dominant constraint is the effective spatial separation between home and workplace. Although precise user-level coordinates are not available, we approximate locations using the geographic coordinates of the serving towers. This approximation is appropriate in our context, as our analysis focuses on prolonged stays at inferred home and work locations rather than on fine-grained movements or interactions with specific transport infrastructures.

For each home–work pair, we compute the Cartesian distance between the corresponding home and work towers and use this value as a proxy for commuting distance. The resulting distribution of commuting distances (reported in the Supplementary Material) is show long tails~\cite{gonzalez2008understanding,brockmann2006scaling,barbosa2018human}.

Based on this distribution, we partition the commuting population into three quantiles containing equal numbers of individuals. The first quantile includes commutes shorter than approximately $2.5$km, the second spans distances between $2.5$km and $5$km, and the third comprises longer commutes exceeding $5$km, with a maximum observed distance of approximately $30$km.\\

\textbf{Administrative Data.} The Bogotá municipal government provides detailed information on the administrative division of the city. Under the current five-year planning framework ($2025-2030$), the administrative structure was revised to ensure that each unit contains a roughly comparable population size, similar socioeconomic conditions, and homogeneous territorial characteristics. As a result, the city is divided into $33$ administrative zones known as Unidades de Planeamiento Local (UPLs, local planning units).

Each UPL is further characterized by a dominant territorial vocation, classified as urban, rural, or mixed urban–rural. This classification reflects the pronounced spatial heterogeneity of Bogotá, which extends beyond socioeconomic differences to include substantial variation in natural and territorial features, with highly urbanized central areas and peripheral zones characterized by mountainous terrain and forested land.

To jointly account for spatial and socioeconomic heterogeneity, UPLs are further subdivided according to socioeconomic strata. This procedure yields a total of $57$ spatial units, each uniquely defined by its administrative boundary, income level, and territorial vocation. This aggregation strategy mitigates issues of low representativeness in sparsely populated or predominantly rural areas while preserving meaningful distinctions in income and land-use characteristics.

Incorporating the administrative level enables spatial aggregation of commuting behavior. In particular, for each origin–destination administrative pair, commuting distance is computed as a weighted mean, where weights correspond to the number of commuters residing in the origin area. This approach allows us to characterize mobility patterns at an intermediate spatial scale that balances geographic resolution and statistical robustness.

\textbf{Digital infrastructure.}
We use fixed broadband Internet speed as a proxy for the quality of local digital infrastructure. To this end, we rely on geolocated Speedtest measurements provided by Ookla, which report download and upload speeds for fixed broadband connections in megabits per second.

For each of the $57$ administrative zones defined above, we compute the median fixed broadband download speed using all available Speedtest observations within the zone. This aggregation is performed separately for each of the three study periods, allowing Internet connectivity measures to be aligned temporally with mobility data. We focus on fixed broadband rather than mobile connectivity, as our analysis concerns prolonged stays at inferred home and workplace locations, for which fixed Internet access is more representative of effective connectivity conditions.

Each commuter is assigned two Internet proxy speed values: one corresponding to the home area and one to the workplace area. These assignments allow us to include digital infrastructure as a covariate in the analysis of commuting behavior. In addition to absolute home Internet speed, we construct a relative connectivity measure defined as the logarithmic ratio between home and workplace download speeds, $log(speed_h/speed_w)$ capturing potential asymmetries in digital access between residential and employment locations.

Speedtest-based measurements are subject to known limitations, including self-selection biases and device- or network-specific effects. However, aggregation at the level of relatively large administrative zones mitigates the influence of idiosyncratic measurements and local artifacts. Ookla Speedtest data have been widely adopted in the literature and by public institutions as a standard proxy for broadband performance~\cite{gozzi2023adoption,gozzi2024bridging,ford2021form,feamster2020measuring}.

\subsection{Segregation}

To quantify income-based segregation at workplaces, we adopt a segregation metric introduced in~\cite{moro2021mobility}. Its generalized form can be expressed as:
\begin{align}
     S_\alpha = \frac{n}{2(n-1)}\sum_{q=1}^{n}\left|\tau_{q\alpha}-\frac{1}{n}\right|
\end{align}
Where $\tau_{q\alpha}$ is the fraction of the population labeled as $q$ (i.e., in our case, low, medium, high income) in a specific location $\alpha$. The metric is designed to capture the extent to which individuals from different socioeconomic groups are unevenly distributed across locations, while accounting for differences in group sizes and overall population composition. The normalization factor makes segregation $S_{\alpha}$ equal to $1$ in case of total segregation (i.e., only one group in $\alpha$) and zero when the the group are equally present in that place. This approach provides a normalized index that is comparable across places and time periods, making it particularly suitable for analyses of segregation dynamics. 

In our case the groups $q$ are three. Hence, the previous equation becomes:
\begin{align}
     S_\alpha = \frac{3}{4}\sum_{q\in SES}\left|\tau_{q\alpha}-\frac{1}{3}\right|
\end{align}
In our box-plots in Fig.~\ref{fig:panel1}C (left panel)  the locations are the work towers, that are then grouped by income in each period. The variation in Fig.~\ref{fig:panel1}C (right panel) is evaluated as a ratio over the median of the baseline period $2019$ and $2020$ up to the announcement of restrictions.

\subsection{Regression analysis}

To identify factors associated with changes in commuting behavior and their implications for spatial inequality, we estimate both univariate (reported in the Supplementary Materials) and multivariate regression models using ordinary least squares (OLS). The dependent variable captures changes in commuting behavior across time by comparing period-specific outcomes to the pre-pandemic baseline.

Each observation in the regression corresponds to one of the $56$ combinations of residential UPL, workplace UPL, and income group introduced in the Results section. For each such combination and for each period, commuting behavior is summarized by the median non-commuting fraction computed across all available working days.

The set of explanatory variables used in the multivariate regression includes both numeric and categorical covariates. Numeric variables capture structural and infrastructural characteristics, such as residential population size, commuting distance, and indicators of digital infrastructure quality. In particular, we include the median fixed broadband download speed in the residential area, as well as the logarithmic ratio between home and workplace internet speeds, capturing relative connectivity conditions. Categorical variables describe socioeconomic status, period, and territorial vocation.

All numeric predictors, as well as the dependent variable, are standardized to zero mean and unit variance prior to estimation. This transformation allows regression coefficients to be compared on a common scale and prevents variables with larger numerical ranges from disproportionately influencing the estimation. Categorical variables are included using dummy encoding, with explicitly defined reference categories.

As an initial step, we estimate univariate regression models in which each predictor is considered separately (see the Supplementary Materials for details). These models provide a descriptive assessment of the marginal association between each covariate and the outcome of interest, without accounting for potential confounding effects. For categorical predictors, all category-specific coefficients are retained except for the reference level.

We then estimate a multivariate regression model including all covariates. This specification allows us to assess the association of each variable with commuting changes while controlling for socioeconomic, spatial, and infrastructural factors. In this framework, coefficients represent conditional associations rather than causal effects.

To quantify uncertainty and improve robustness, we rely on non-parametric bootstrap resampling rather than analytical standard errors. For both univariate and multivariate models, we generate $500$ bootstrap samples by resampling observations with replacement and re-estimating the model on each replicate. The resulting empirical distributions of the coefficients are used to compute confidence intervals reported in the Results section. This approach is well suited to our setting, as the dependent variables are derived from aggregated mobility measures and the observations correspond to spatially structured home--work area combinations, which may violate standard OLS assumptions.

For multivariate regressions, we additionally record the coefficient of determination (i.e., $R^2$) for each bootstrap replicate to assess model fit. The $R^2$ values reported in the Results section correspond to the median of the bootstrap distribution.

While OLS provides a transparent and interpretable framework, all regression results should be interpreted as associational rather than causal. Moreover, although bootstrapping improves robustness to distributional assumptions, it does not correct for potential biases arising from omitted variables or measurement error. For these reasons, regression results are complemented by additional robustness checks and other models (decision trees regressors) presented in the Supplementary Materials.

\section{Acknowledgments}
All authors thank Ookla, The World Bank and the Development Data Partnership. All authors thank James Carroll, and Katherine Macdonald for their support and review.

\section{Author contributions}
All authors designed the study, wrote and approved the manuscript. C.M. and N.G. performed the analyses. L.F. mined and extracted the mobility data.

\section{Funding}
The research is partially funded by The World Bank Digital Development Partnership. The findings, interpretations, and conclusions expressed in this paper are entirely those of the authors. They do not necessarily represent the views of the International Bank for Reconstruction and Development/World Bank and its affiliated organizations, or those of the Executive Directors of the World Bank or the governments they represent.

C.M., L.F. and N.G. acknowledge support from the Lagrange Project of the Institute for Scientific Interchange Foundation (ISI Foundation) funded by Fondazione Cassa di Risparmio di Torino (Fondazione CRT). L.F. also acknowledges support from the Fondo de Investigación y Desarrollo en Salud, Fonis, Project SA24I0124.

\section{Data availability}
All the dataset used are publicly available and referenced within the text with the exception of the internet Speedtest measurements from Ookla and the raw mobility data. The Ookla dataset is proprietary and cannot be shared publicly. The raw mobility data cannot be shared due to privacy concerns. Only aggregated mobility patterns across geographical units have been used for the results presented here. The use of mobility data in this study was deemed exempt (Folio: 24022026LAF) by the Comité de Ética Institucional en Investigación at Universidad del Desarrollo.

\bibliographystyle{unsrtnat}
\bibliography{bibl}

\section*{Supplementary materials}

\subsection*{Population validation using high-resolution density data}

To assess its representativeness, we compare the population data derived from the mobile phone data against an independent source of population distribution at similar spatial scale. To achieve this, we used the High Resolution Population Density Maps provided by Meta Data for Good~\cite{density_maps}. This dataset is based on satellite imagery and census data and provide gridded population estimates with a maximum spatial resolution of approximately $30$ meters. The spatial resolution of these tiles is comparable to that of antennas, allowing us to define population estimates within administrative boundaries with a level of spatial detail consistent with our mobility data. We performed a spatial aggregation of the population tiles within the boundaries of each administrative unit (i.e., \textit{localidad}) of Bogotá. In parallel, using our mobile phone dataset, we computed for each unit the number of users associated with home antennas located within the same administrative boundaries. 

\begin{figure}[ht]
    \centering
    \includegraphics[width=1\textwidth]{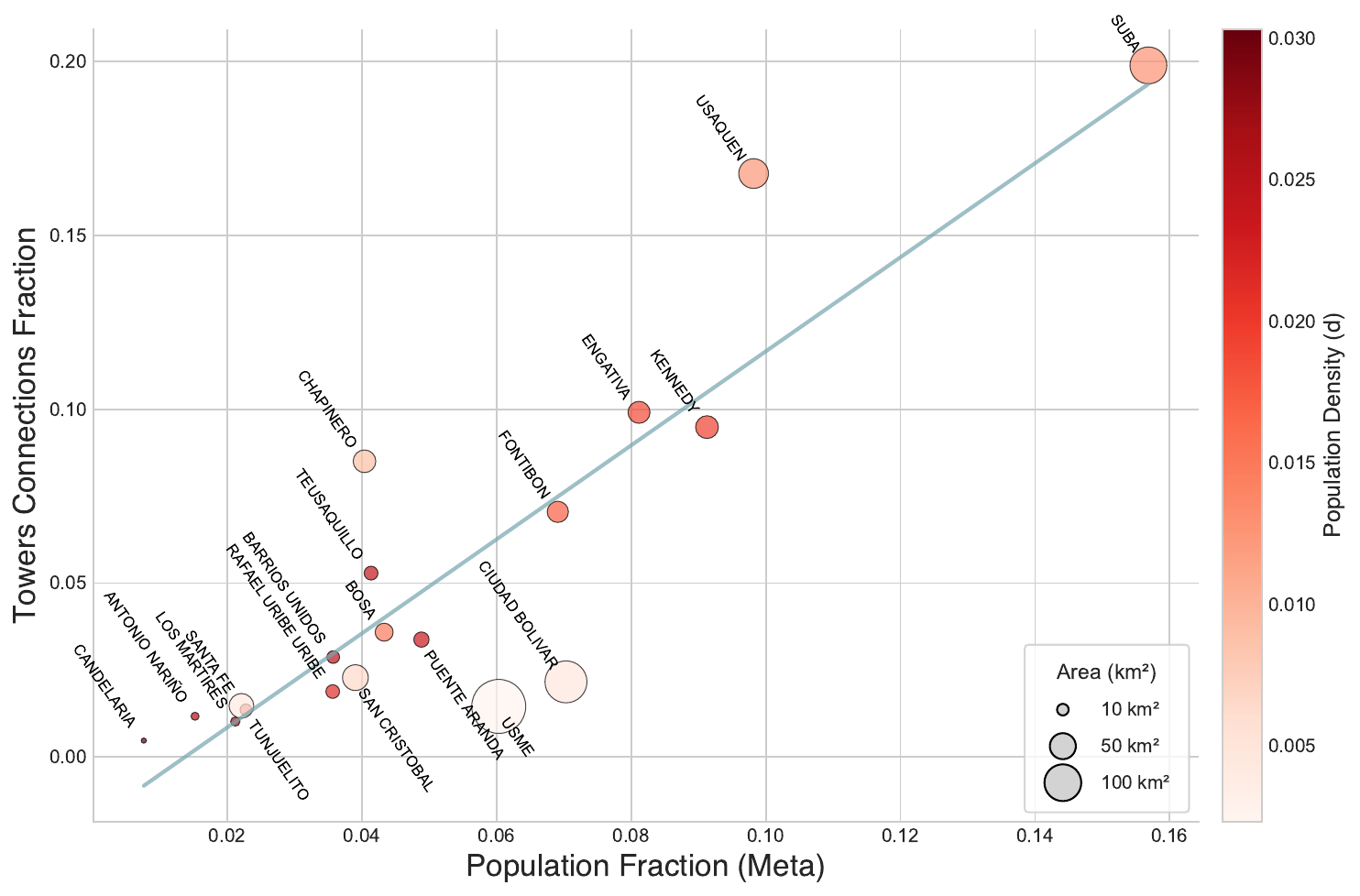}
    \caption{Scatter plot comparing population fractions estimated from Meta Data for Good high-resolution density maps (x-axis) and fractions of users associated with home antennas (y-axis), aggregated by administrative unit. Marker size represents unit area ($km^2$), and color indicates population density. The Pearson correlation ($0.88$) confirms strong agreement between the two independent population proxies.}
    \label{fig:scatter_corr}
\end{figure}

The comparison between the two population estimates is provided in Fig.~\ref{fig:scatter_corr}. The Pearson correlation coefficient between the two population estimates is $0.88$, indicating a strong, positive and significant correlation between them. This high correlation supports the use of the number of users associated with home antennas as a proxy for population at the administrative unit level in the main analysis.

\subsection*{Exploring the role of commuting distance}

This section complements the main analysis by providing additional detail on how commuting distance is defined and how it interacts with socioeconomic stratification.

Commuting distance is approximated simply as the Cartesian distance between the geographic coordinates of the towers inferred as home and work locations for each individual. This measure does not capture route-specific distances, but provides a consistent and scalable proxy for effective spatial separation between home and workplace across the entire metropolitan area. Based on the empirical distribution of distances, we partition commuting flows into three groups (based on the terciles of the distribution) containing an equal shares of commuters: short-distance commutes (below approximately $2.5$ km), medium-distance commutes ($2.5$–$5$ km), and long-distance commutes (above $5$ km).

\begin{figure}[!ht]
    \centering
    \includegraphics[width=1\textwidth]{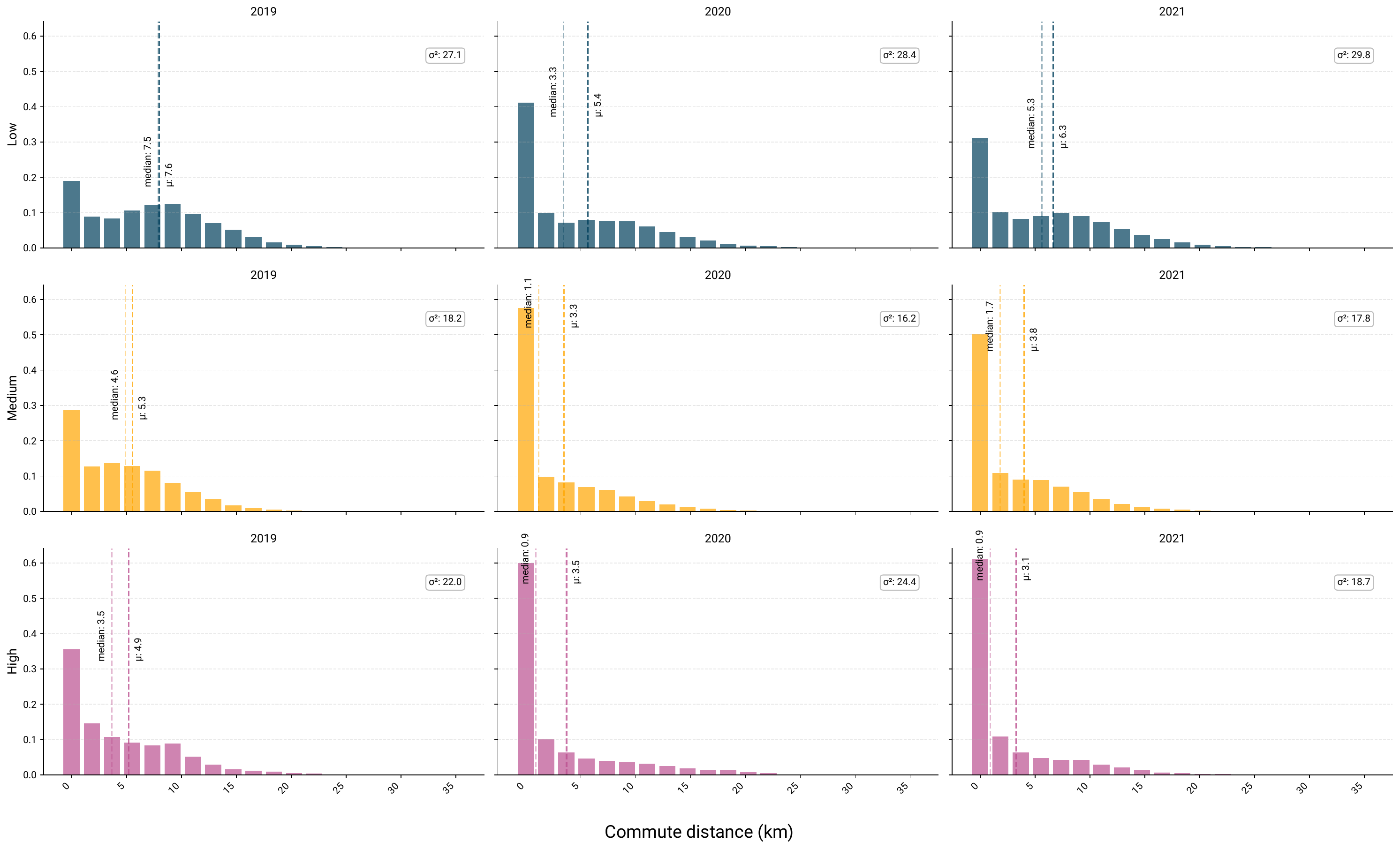}
    \caption{Distribution of the residential population traveled distance to work location per residential SES and year.}
    \label{fig:distance_distribution}
\end{figure}

Fig.~\ref{fig:distance_distribution} shows the distribution of commuting distances by residential socioeconomic group and year. Across all groups and periods, the distributions are strongly right-skewed and exhibit heavy-tailed behavior, consistent with established empirical regularities in human mobility. During the restriction period ($2020$), the distributions contract markedly, with a reduction in long-distance commuting across all income groups, followed by a partial re-expansion in $2021$. Higher-income groups consistently display shorter typical commuting distances.

\begin{figure}[!ht]
    \centering
    \includegraphics[width=1\textwidth]{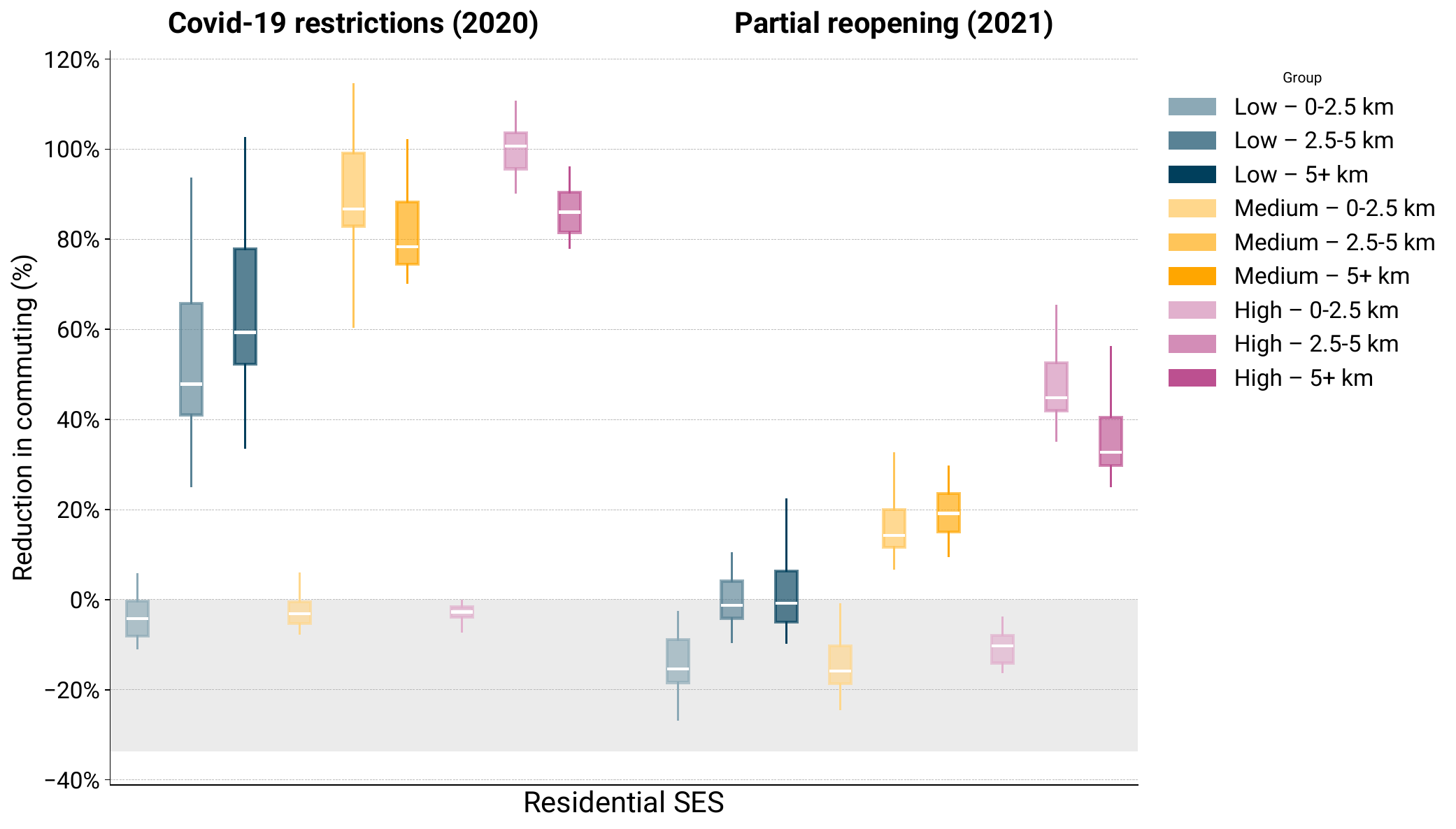}
    \caption{Reduction in commuting relative to the $2019$ baseline, stratified by \textbf{residential SES} during COVID-19 restrictions ($2020$) and partial reopening ($2021$) for traveled distance to work.}
    \label{fig:residential_distance}
\end{figure}

\begin{figure}[!ht]
    \centering
    \includegraphics[width=1\textwidth]{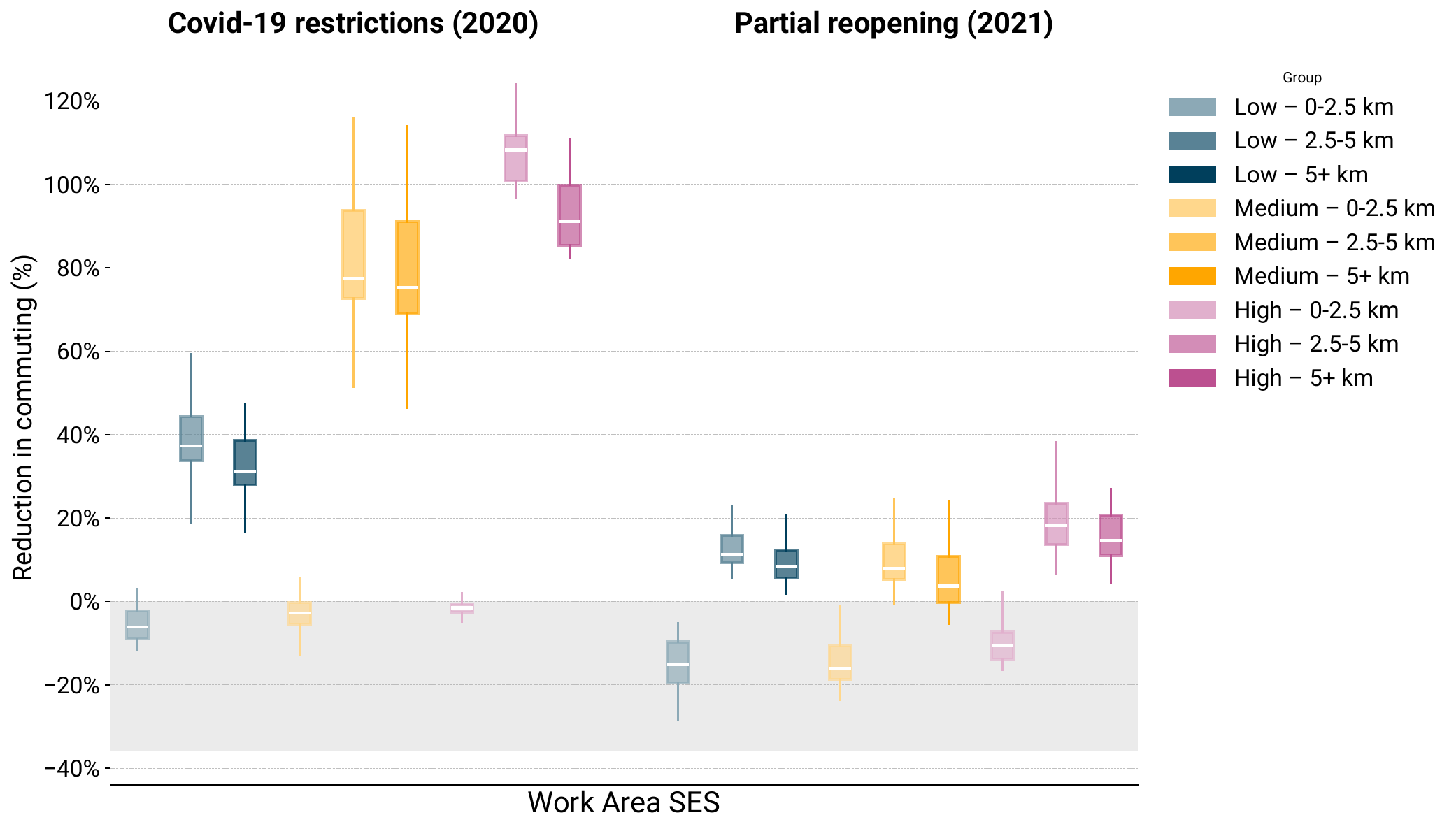}
    \caption{Reduction in commuting relative to the $2019$ baseline, stratified by \textbf{work area SES} during COVID-19 restrictions ($2020$) and partial reopening ($2021$) for traveled distance to work.}
    \label{fig:work_distance}
\end{figure}

Fig.~\ref{fig:residential_distance} Fig.~\ref{fig:work_distance} examine how changes in mobility vary by commuting distance and SES, from two complementary perspectives. Fig.~\ref{fig:residential_distance} reports boxplots of commuting reduction stratified by residential SES, capturing mobility changes from the point of view of workers (i.e., grouping individuals according to the SES of their home areas). Fig.~\ref{fig:work_distance} instead stratifies results by work-area SES, adopting a perspective that reflects how different types of employment locations experienced changes in the presence of workers from different distances.

According to both perspectives, longer commuting distances are associated with larger decrease in commuting during the restriction period ($2020$), indicating that long-distance commutes were more likely to be suspended. However, this distance gradient is markedly stronger for medium and high-income groups, while low-income commuters exhibit weaker distance-dependent variation, consistent with more limited flexibility to reduce travel even for longer commutes. The work-area perspective further shows that reductions in commuting were not evenly distributed across employment locations and commuting distances, with high-SES work areas exhibiting larger contractions in long-distance inflows.

Together, these results motivate the inclusion of commuting distance as a key explanatory variable in the regression analysis and clarify how distance-dependent effects differ when viewed from residential versus workplace vantage points.

\subsection*{Correlation and partial correlation analysis of commuting behavior change}

To assess relationships among commuting reduction and its possible predictors, we computed Pearson correlations and pairwise partial correlations. This analysis aims to reveal the degree of association between each predictor and the variation in commuting behavior but also to identify potential multi-collinearity among predictors that could affect regression estimates. The partial correlation coefficients further clarify whether observed correlations are direct or confounded by third variables. All correlation computations were performed using the \textit{pingouin} Python package~\cite{vallat2018pingouin}.

\subsubsection*{Correlations among predictors}

\begin{figure}[ht]
    \centering
    \includegraphics[width=0.85\textwidth]{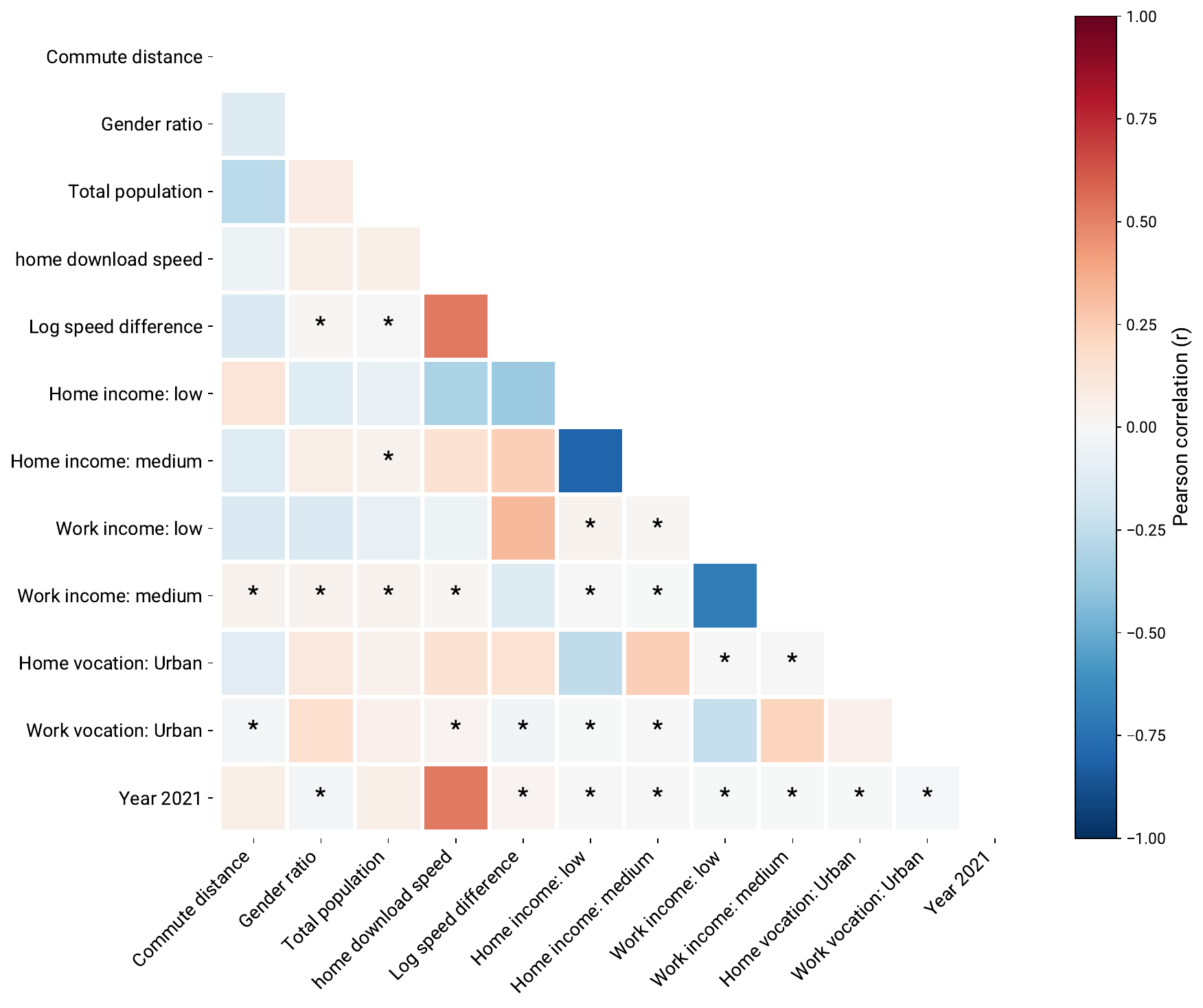}
    \caption{Pairwise Pearson correlations among predictors. Lower triangle shows correlation coefficients. Color intensity and hue indicate magnitude and direction (red = positive, blue = negative). Asterisks mark non-significant correlations ($p > 0.05$). Negative correlations between dummies of the same categorical variable (e.g., home income: low vs. medium) are structural artifacts of one-hot encoding and should not be interpreted as meaningful associations.}
    \label{fig:predictor_corr}
\end{figure}

We first computed the full pairwise correlation matrix among all predictors (Fig.~\ref{fig:predictor_corr}). This matrix includes correlations between continuous variables, binary dummy variables derived from categorical features, and mixed pairs. The interpretation of these correlations depends on the variable types involved:

\begin{itemize}
    \item \textbf{Continuous $\times$ Continuous}: Standard Pearson correlation coefficient, measuring the strength and direction of linear association between two numerical variables (e.g., commuting distance and total population).
    \item \textbf{Binary $\times$ Binary}: Also computed as Pearson correlation, but mathematically equivalent to the $\phi$ coefficient, which  measures association between two dichotomous variables. A positive correlation indicates that observations in one category are more likely to also be in the other category (e.g., urban home areas residents tend to have urban work areas). We note how when a categorical variable with three levels (e.g., income: low, medium, high) is one-hot encoded with one category dropped as reference, each resulting dummy variable is binary and correlations involving these dummies are mathematically binary$\times$binary. However, the interpretation must account for the reference category: a correlation between a dummy and another variable compares that specific category against all other categories pooled together (medium and high combined), not against the reference alone. Additionally, when a categorical variable with three levels (e.g., income: low, medium, high) is one-hot encoded with one category dropped as reference, the resulting dummy variables are mechanically negatively correlated with each other. These correlations are artifacts of the encoding scheme and do not reflect meaningful associations.
    \item \textbf{Binary $\times$ Continuous}: Pearson correlation in this context is equivalent to the point-biserial correlation, measuring the difference in means of the continuous variable between the two groups defined by the binary variable.
\end{itemize}

Pairwise correlations are shown in Fig.~\ref{fig:predictor_corr}. Statistical significance was assessed using a two-tailed test of the null hypothesis. Non-significant correlations with $p > 0.05$ are marked with an asterisk. The correlation matrix reveals generally low to moderate pairwise associations among predictors, with no evidence of severe multi-collinearity. The only relatively strong correlations arise mechanically among dummy variables derived from the same categorical factor and between download speed and log speed difference, which capture closely related constructs. Consistent with the main results, the $2021$ indicator is positively correlated with home download speed, reflecting the expected increase in broadband performance over time.

\subsubsection*{Simple and partial correlations with the commuting reduction}

\begin{figure}[ht]
    \centering
    \includegraphics[width=0.95\textwidth]{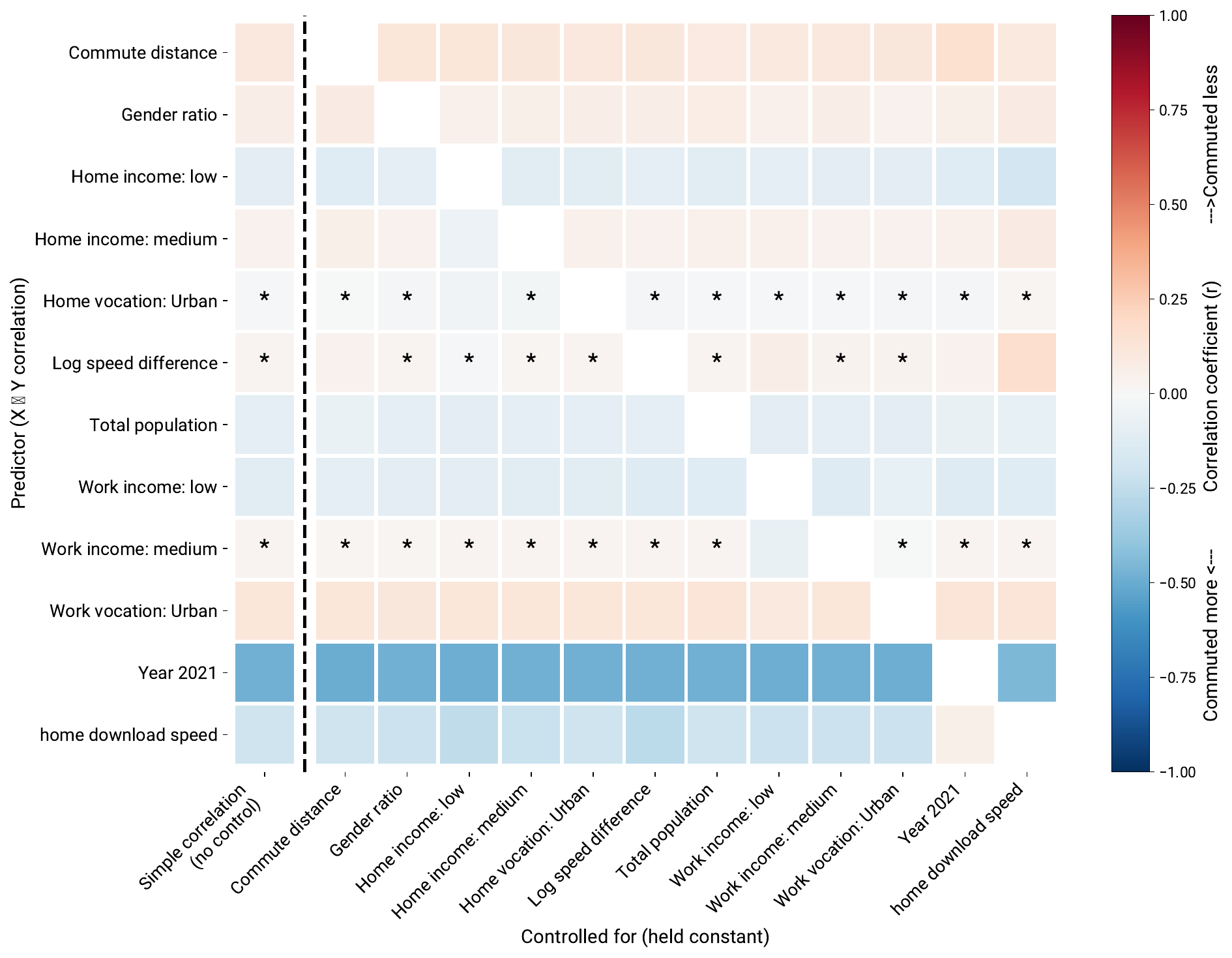}
    \caption{Simple and partial correlations with variation in commuting behavior. The first column (separated by a dashed line) shows simple Pearson correlations between each predictor and the outcome. Subsequent columns display partial correlations: correlation between each predictor (row) and the outcome after controlling for the variable listed in the column header. Asterisks indicate non-significant associations ($p > 0.05$). Comparing the first column to the partial correlation columns reveals which predictors mediate or confound the relationships observed in simple bivariate analysis.}
    \label{fig:partial_corr}
\end{figure}

Fig.~\ref{fig:partial_corr} displays both simple and pairwise partial correlations between each predictor and the variation in commuting behavior relative to baseline. The first column shows the simple Pearson correlation of each predictor with the outcome, ignoring all other variables. The remaining columns display partial correlations: the correlation between each predictor (row) and the outcome after linearly removing the effect of a single control variable (column). Partial correlation measures the strength of association between two variables while holding a third variable constant. In our pairwise partial correlation matrix, we compute the partial correlation between each predictor and the outcome while controlling for each of the other predictors one at a time, allowing us to assess which variables explain away or mediate the associations observed in the simple correlations. The partial correlation analysis anticipates the multivariate regression results. In general, variables that are positively correlated with commuting reduction tend to display positive regression coefficients (and vice versa), reflecting the direction of the conditional association once other factors are jointly considered. In particular, the period indicator (2021) emerges as the strongest negative association in both the correlation matrix and the regression model, while work vocation shows the strongest positive association. Importantly, these effects remain substantial even when mediated by other covariates in the partial correlation exercise, confirming their structural robustness in the full multivariate specification.

\subsection*{Sensitivity of main findings to work-day definition}

In the results discussed in the main text, work locations are inferred using weekday connections observed between $09:00$ and $17:00$ (Bogotá local time). This time window is commonly adopted to capture standard business hours. However, working schedules in Latin American cities can start earlier, particularly in sectors such as retail, manufacturing, and public services. To assess the robustness of our results to the definition of working hours, we repeated the full pipeline using alternative morning start times of $08:00$ and $07:00$.

For each alternative window (i.e., $07:00–17:00$ and $08:00–17:00$), we re-identified work locations, reconstructed commuting flows, recomputed the non-commuting fraction and its variation relative to the pre-pandemic baseline, and repeated all descriptive and regression analyses.

\begin{figure}[!ht]
    \centering
    \includegraphics[width=0.85\textwidth]{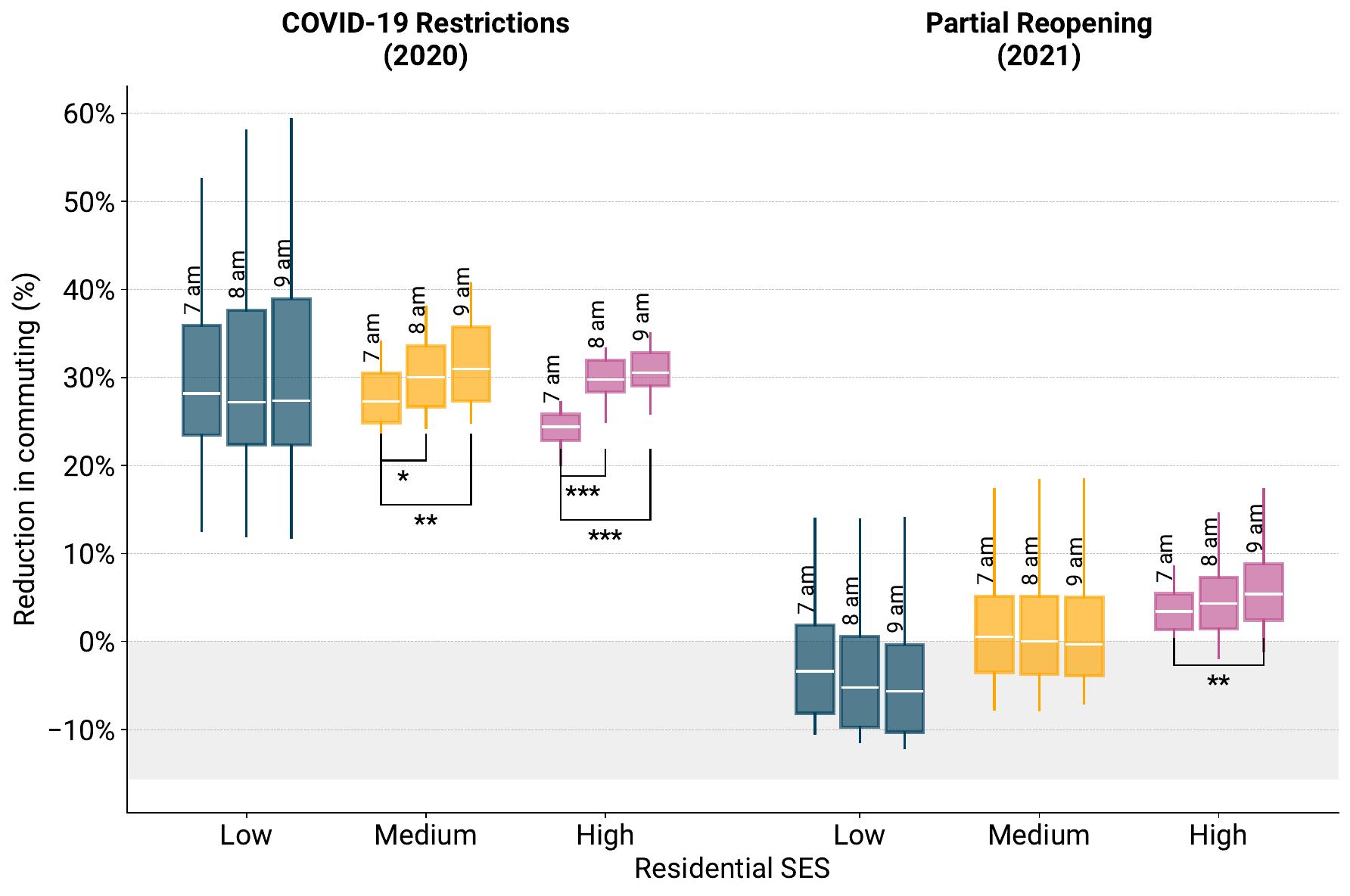}
    \caption{Reduction in commuting relative to the $2019$ baseline, stratified by \textbf{residential SES} during COVID-19 restrictions ($2020$) and partial reopening ($2021$) for the three different morning times. Asterisks indicate statistically significant differences between working hours definitions within the same SES groups and period measured via Mann-Whitney U test. The number of asterisks reflecting the magnitude of the p-value:  *** ($p \leq 0.001$), ** ($0.001 \leq p < 0.02$), * ($0.02< p <0.05$), blank otherwise.}
    \label{fig:morning_times_home}
\end{figure}

\begin{figure}[!ht]
    \centering
    \includegraphics[width=0.85\textwidth]{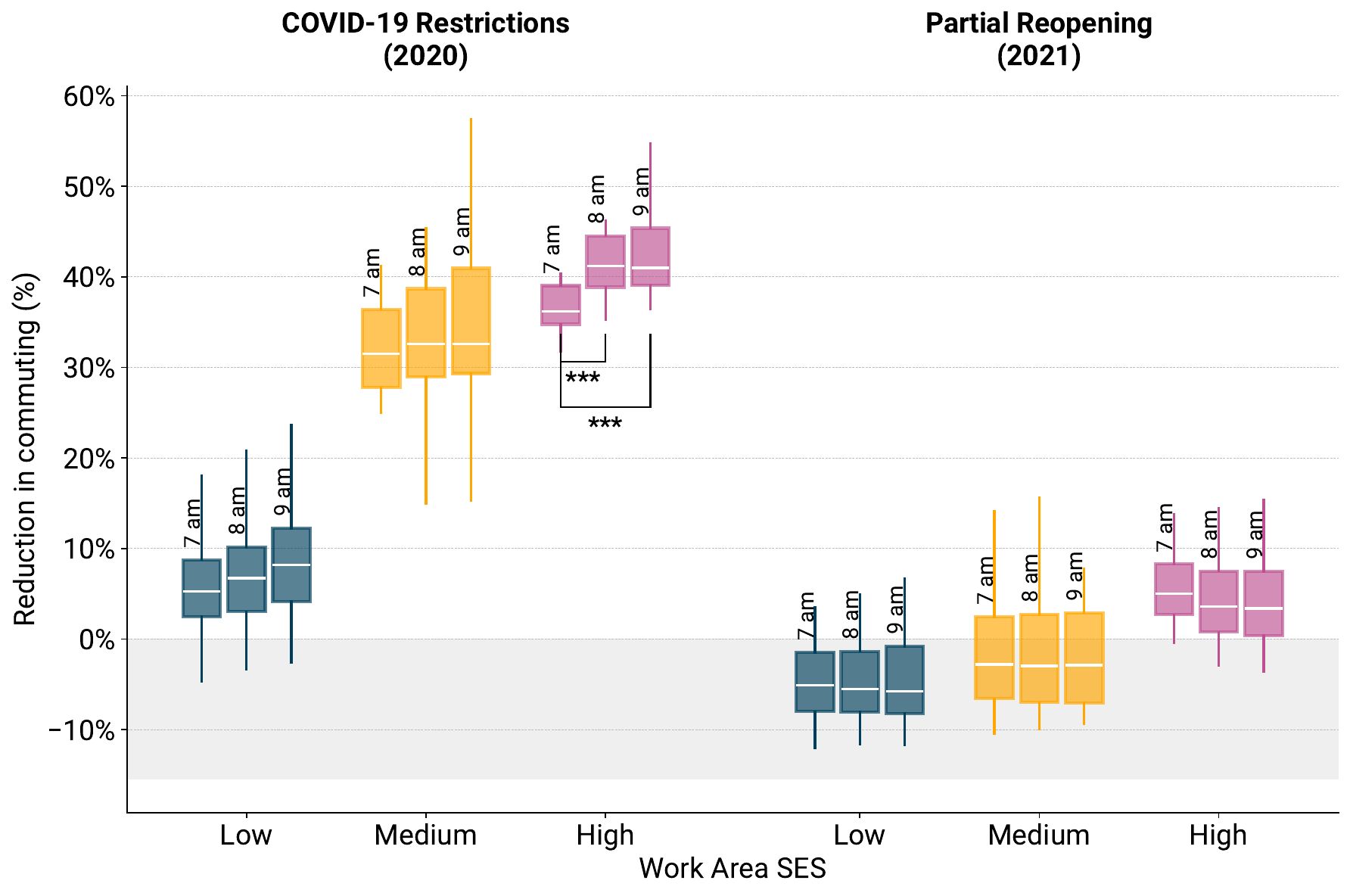}
    \caption{Reduction in commuting relative to the $2019$ baseline, stratified by \textbf{work area SES} during COVID-19 restrictions ($2020$) and partial reopening ($2021$) for the three different morning times. Asterisks indicate statistically significant differences between working hours definitions within the same SES groups and period measured via Mann-Whitney U test. The number of asterisks reflecting the magnitude of the p-value:  *** ($p \leq 0.001$), ** ($0.001 \leq p < 0.02$), * ($0.02< p <0.05$), blank otherwise.}
    \label{fig:morning_times_work}
\end{figure}

Fig.~\ref{fig:morning_times_home} and Fig.~\ref{fig:morning_times_work} compares the reduction in commuting by, respectively, residential and workplace SES across the three working-hour definitions. The overall patterns remain stable. In all specifications, commuting reductions during the restriction period are broadly similar across SES groups, while differences emerge primarily in the reopening phase, with lower-SES groups showing a stronger rebound. The relative ordering of SES groups is preserved across working hours definitions. To formally test whether alternative working hours definitions produce statistically different commuting reductions, we performed non-parametric Mann–Whitney tests comparing the distributions obtained under the working hours window starting at $09:00$ with those obtained for windows starting at the $08:00$ and $07:00$. Statistically significant differences are marked in figure with asterisks. Across SES categories and periods, we find no systematic evidence of substantial distributional shifts. While statistical differences in medians are occasionally detected, effect sizes remain small relative to the overall variability of the data.

Before estimating the multivariate specification, we first examine the relationship between commuting reductions and each explanatory variable separately through a set of univariate regressions. This step provides a preliminary assessment of the direction and magnitude of associations, while allowing us to verify whether the patterns observed in the descriptive analysis persist when variables are considered individually.

\begin{figure}[ht]
    \centering
    \includegraphics[width=1\textwidth]{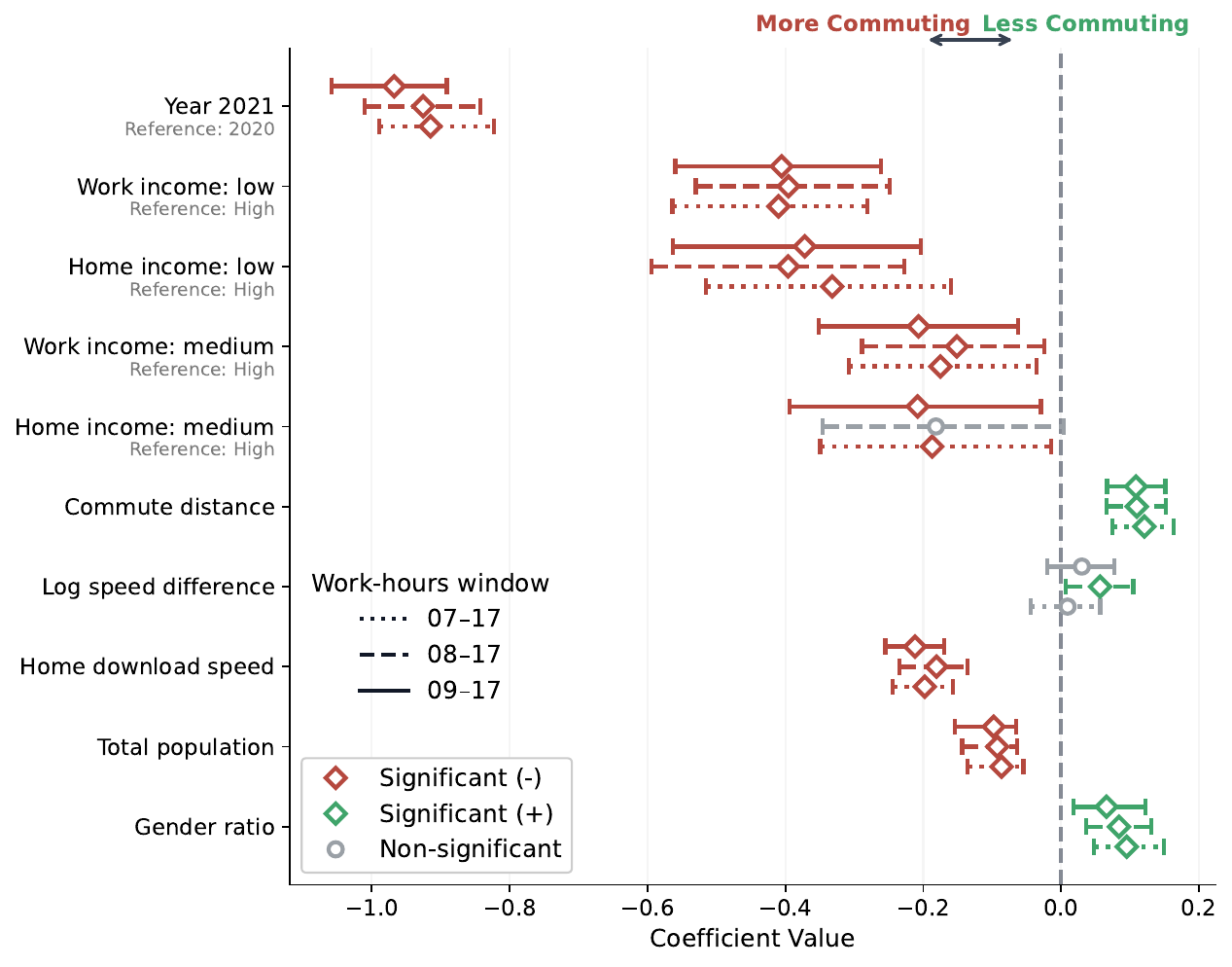}
    \caption{{Univariate regression coefficients for different working-hours window definitions with $95\%$ confidence intervals. Red/green indicates statistically significant effects (negative and positive) at the $5\%$ confidence level; gray indicates non-significant effects.}}
    \label{fig:regression_univariate_morning_times}
\end{figure}

Fig.~\ref{fig:regression_univariate_morning_times} reports the estimated coefficients and $95\%$ confidence intervals obtained under the three alternative working-hour definitions via univariate regressions. Across specifications, the sign and relative magnitude of most coefficients are consistent with the correlation analysis and stable across different working hours definition. Variables related to socioeconomic stratification, commuting distance, and connectivity conditions display consistent associations with commuting reductions across working-hour windows, suggesting that the choice of morning start time does not substantially alter the underlying relationships.

\begin{figure}[ht]
    \centering
    \includegraphics[width=1\textwidth]{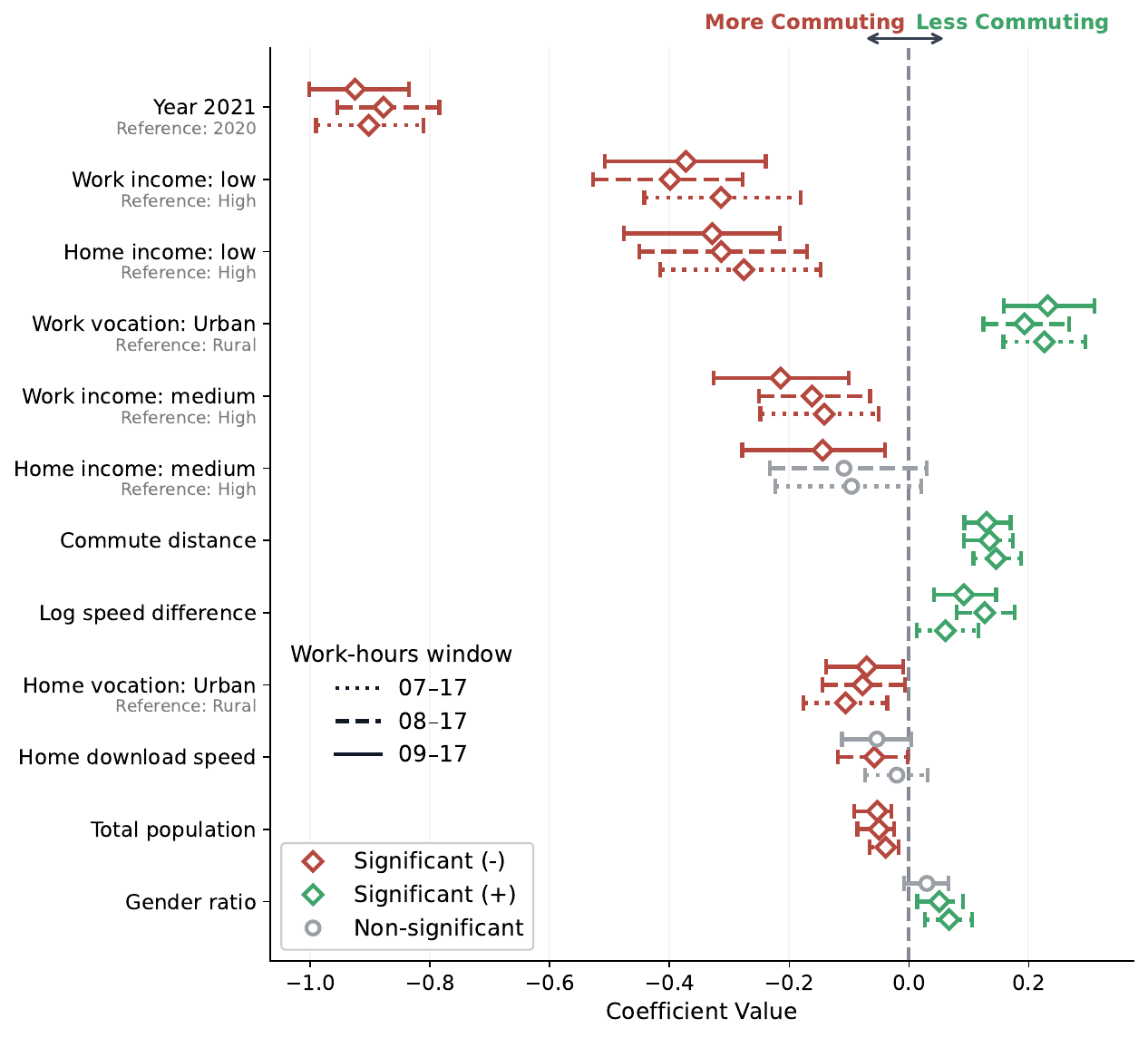}
    \caption{Multivariate regression coefficients for different working hours window definition with $95\%$ confidence intervals, ranked by absolute magnitude. Red/green indicates significant effects (negative and positive) at the $5\%$ confidence level; gray indicates non-significant effects.}
    \label{fig:regression_morning_times}
\end{figure}

We further compare the multivariate regression results obtained under the three working hours window definitions. Fig.~\ref{fig:regression_morning_times} shows regression coefficients for all covariates obtained defining the commuting reduction with respect to the three alternative working hours definitions. The direction and statistical significance of the main predictors — including period, socioeconomic stratification, commuting distance, and relative internet speed — remain consistent across specifications. Minor fluctuations are observed in the magnitude of certain coefficients. The explanatory power of the model is also comparable across definitions, with median $R^2$ values of approximately $0.27$ for the $07:00$ window, $0.28$ for the $08:00$ window, and $0.30$ for the $09:00$ window.

In conclusion, given the consistency of descriptive patterns, statistical tests, and regression coefficients across definitions, we conclude that the results are robust to plausible variations in the start time used for work detection. We retain the $09:00–17:00$ specification in the main text because it aligns more closely with conventional definitions of standard working hours, while minimizing potential contamination from early-morning non-work activities.

\subsection*{Alternative regression analysis}

To assess the robustness of the multivariate linear regression analysis to model choice, we fitted two alternative machine learning models to predict the variation in commuting behavior relative to the pre-pandemic baseline. We consider two tree-based ensemble machine learning models, namely a Random Forest and an AdaBoost regressor. These models were estimated separately for each of the three working hours definitions ($07:00$, $08:00$, and $09:00$). All models were implemented using the Python package \texttt{scikit-learn}~\cite{scikit-learn}.

\subsubsection*{Model setup and hyperparameter search}

The feature set includes the same variables of the multivariate regression presented in the main text. Both models were trained on a $75/25$ train-test split. Hyperparameters were optimized via randomized search over $30$ parameter combinations, evaluated using $5$-fold cross-validation using $R^2$ as performance metric.

For the \textbf{Random Forest}, the search explored the number of trees, maximum tree depth, minimum samples required to split a node or constitute a leaf, and the number of features considered at each split. Out-of-bag scoring was enabled as an additional internal estimate of generalization error, obtained by evaluating each tree on the training observations not included in its bootstrap sample. The optimal hyperparameters sets are reported in Tab.~\ref{tab:best_params}.

For \textbf{AdaBoost}, the base learner was a decision tree regressor. The hyperparameters search covered the number of boosting rounds, the learning rate controlling the contribution of each tree, and the maximum depth of the base trees. Selected hyperparameters are also reported in Tab.~\ref{tab:best_params}.

\begin{table}[ht]
\centering
\caption{Best hyperparameters selected by randomized cross-validated search for Random Forest (RF) and AdaBoost (Ada) across morning-time definitions.}
\label{tab:best_params}
\begin{tabular}{llccccc}
\hline
Window & Model & Trees & Depth & Min split & Min leaf & Features / LR \\
\hline
09:00 & RF  & 200 & 10   & 5  & 2 & $\log_2$ \\
09:00 & Ada & 200 & 4    & -- & -- & LR $= 0.01$ \\
\hline
08:00 & RF  & 200 & None & 2  & 4 & sqrt \\
08:00 & Ada & 100 & 4    & -- & -- & LR $= 0.05$ \\
\hline
07:00 & RF  & 400 & 10   & 10 & 2 & $\log_2$ \\
07:00 & Ada & 200 & 4    & -- & -- & LR $= 0.01$ \\
\hline
\end{tabular}
\end{table}

\subsubsection*{Model evaluation}

Both models were evaluated on the held-out test set using the coefficient of determination ($R^2$) and root mean squared error (RMSE). Performance was computed on both training and test sets to assess overfitting. Results are
reported in Tab.~\ref{tab:model_eval}.

\begin{table}[ht]
\centering
\caption{Train and test performance of Random Forest and AdaBoost regressors
across the three morning-time definitions.}
\label{tab:model_eval}
\begin{tabular}{llcccc}
\hline
Window & Model & Train $R^2$ & Train RMSE & Test $R^2$ & Test RMSE \\
\hline
09:00 & Random Forest & 0.66 & 0.24 & 0.42 & 0.35 \\
09:00 & AdaBoost      & 0.41 & 0.32 & 0.42 & 0.35 \\
\hline
08:00 & Random Forest & 0.61 & 0.27 & 0.40 & 0.34 \\
08:00 & AdaBoost      & 0.46 & 0.32 & 0.35 & 0.35 \\
\hline
07:00 & Random Forest & 0.59 & 0.28 & 0.38 & 0.33 \\
07:00 & AdaBoost      & 0.37 & 0.34 & 0.36 & 0.33 \\
\hline
\end{tabular}
\end{table}

Random Forest consistently shows a larger gap between training and test $R^2$, indicating moderate overfitting despite regularization through the hyperparameter search. AdaBoost achieves more balanced train-test performance. Across all specifications, test $R^2$ values are broadly consistent with those obtained by the linear regression model, suggesting that the nonlinear ensemble models do not uncover substantially different structure in the data beyond what is already captured by the linear specification.

\subsubsection*{Features' importance}

To interpret the contribution of individual features to model predictions, we computed feature importance using two different approaches: SHAP (SHapley Additive exPlanations) values and permutation importance.

\subsubsection*{SHAP values analysis}
SHAP values decompose each prediction into additive contributions from individual features, grounded in cooperative game theory, providing both the direction and magnitude of each feature's effect for every observation~\cite{lundberg2017unified}. For the Random Forest, we used a tree-aware explainer that exploits the model structure to compute exact Shapley values efficiently. For AdaBoost, a model-agnostic kernel-based explainer was used instead, which approximates Shapley values via weighted linear regression over feature coalitions.

Fig.~\ref{fig:shap_importance_all} displays, for each feature, the full distribution of SHAP values across the sample, with color encoding whether the original feature value was high or low. Features are ranked by their mean absolute SHAP value, providing a global measure of importance alongside directional information.

Across all three working hours definitions and both models, the period indicator is by far the most important feature. Commuting distance ranks consistently second, with longer commutes associated with larger reductions in presence at work. Work-area urban vocation also appears among the most important features in all specifications. Income-related variables and Internet speed metrics contribute more modestly but remain directionally
consistent. The overall ranking of features is stable across working hours definitions, reinforcing the robustness of the identified predictors.

\subsubsection*{Permutation importance}

\begin{figure}[ht]
    \centering

    \begin{subfigure}[b]{0.32\textwidth}
        \centering
        \includegraphics[width=\textwidth]{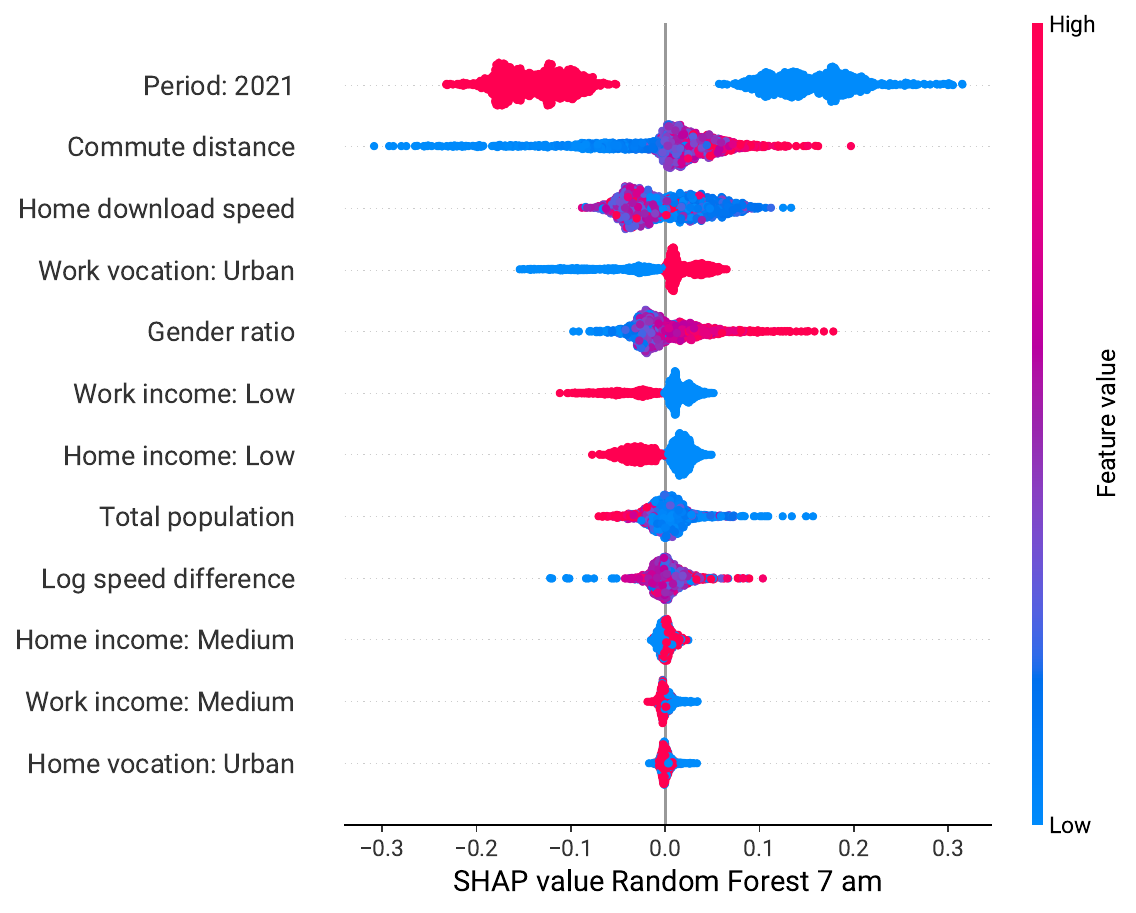}
        \caption{SHAP -- RF, 07:00}
        \label{fig:shap_rf_7}
    \end{subfigure}
    \hfill
    \begin{subfigure}[b]{0.32\textwidth}
        \centering
        \includegraphics[width=\textwidth]{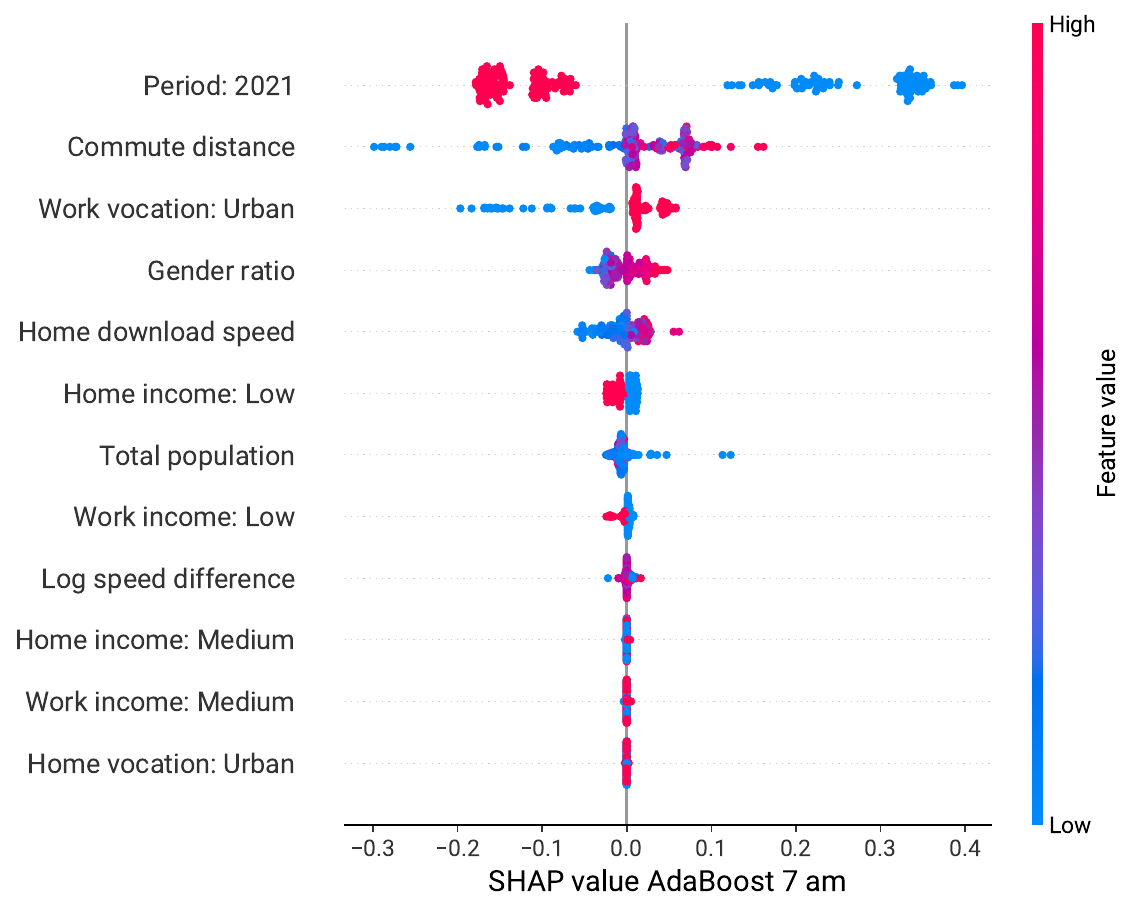}
        \caption{SHAP -- AdaBoost, 07:00}
        \label{fig:shap_ada_7}
    \end{subfigure}
    \hfill
    \begin{subfigure}[b]{0.32\textwidth}
        \centering
        \includegraphics[width=\textwidth]{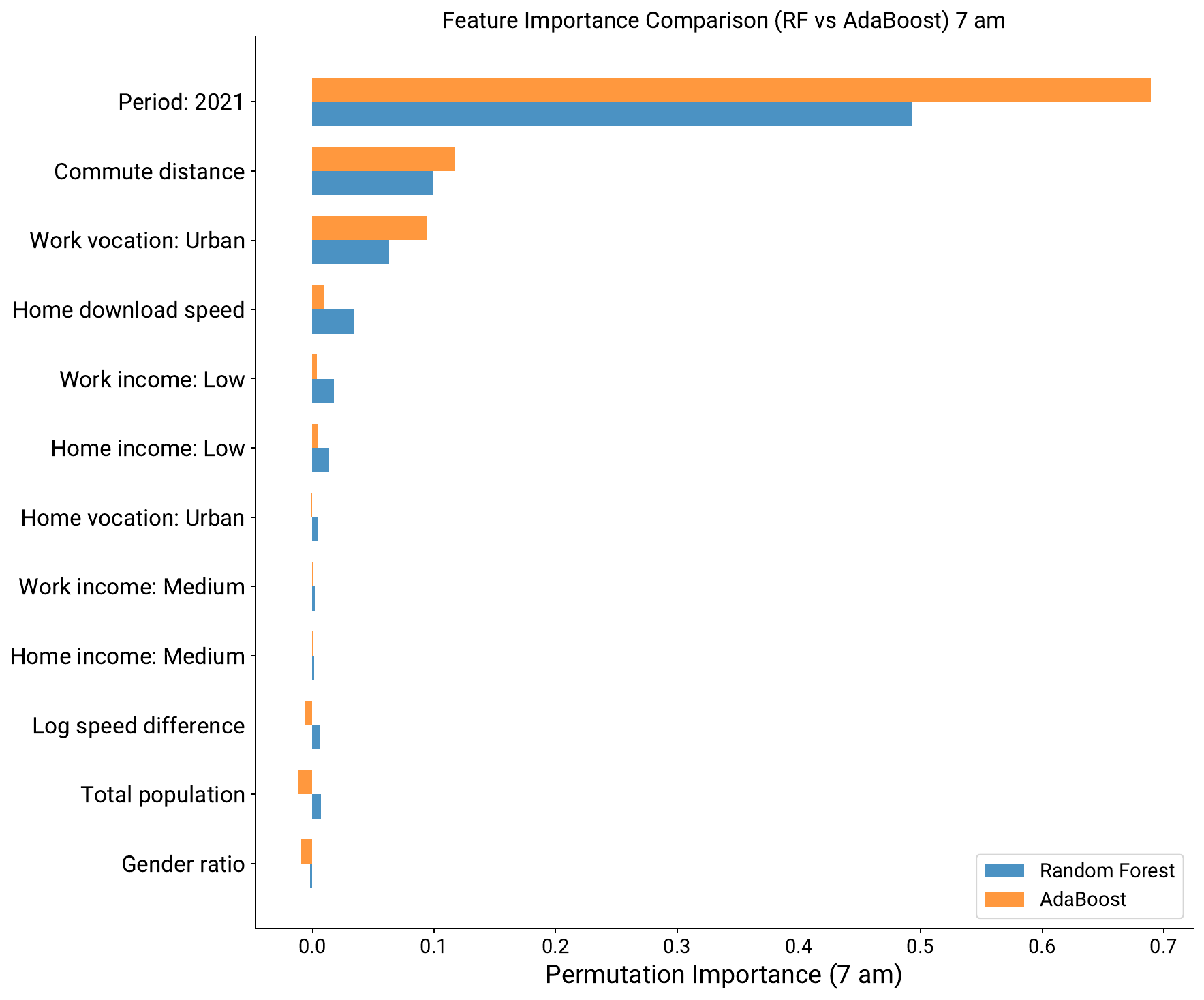}
        \caption{Permutation importance, 07:00}
        \label{fig:imp_7}
    \end{subfigure}

    \vspace{0.5em}

    \begin{subfigure}[b]{0.32\textwidth}
        \centering
        \includegraphics[width=\textwidth]{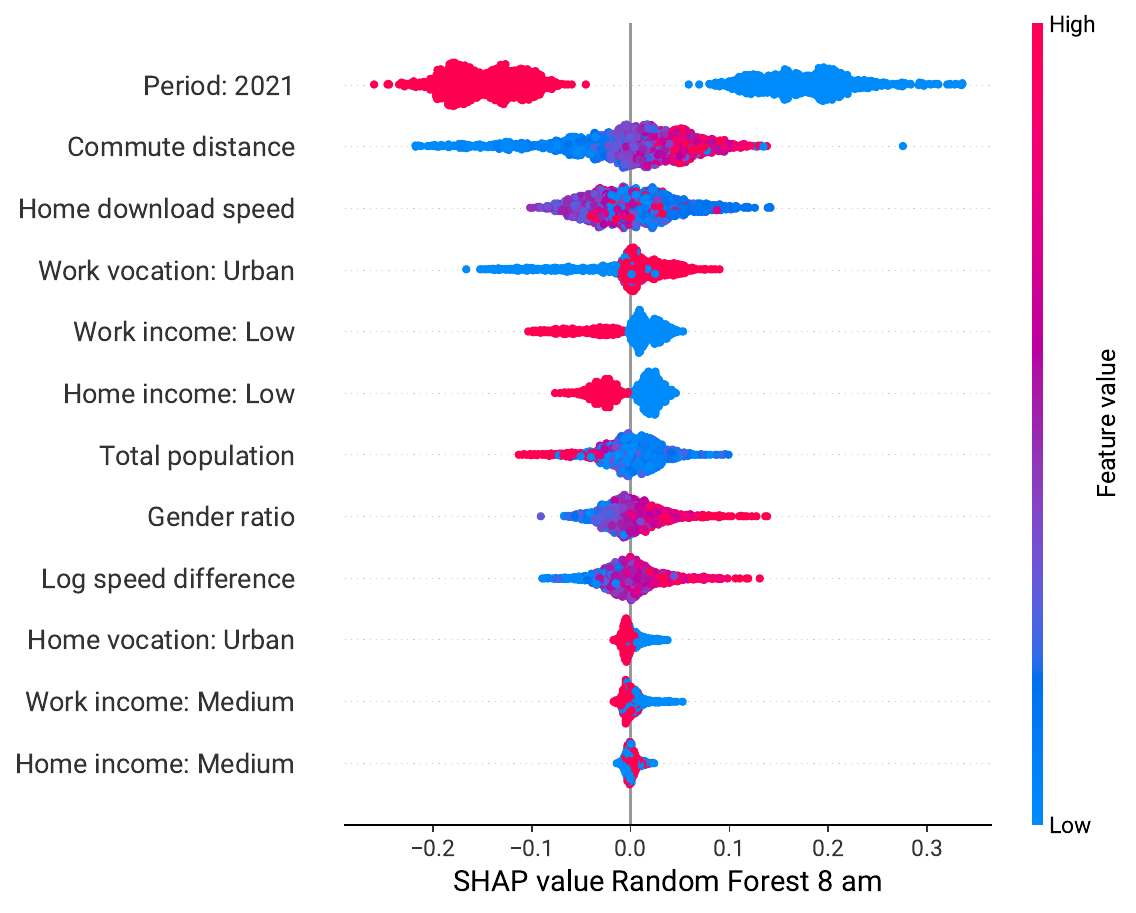}
        \caption{SHAP -- RF, 08:00}
        \label{fig:shap_rf_8}
    \end{subfigure}
    \hfill
    \begin{subfigure}[b]{0.32\textwidth}
        \centering
        \includegraphics[width=\textwidth]{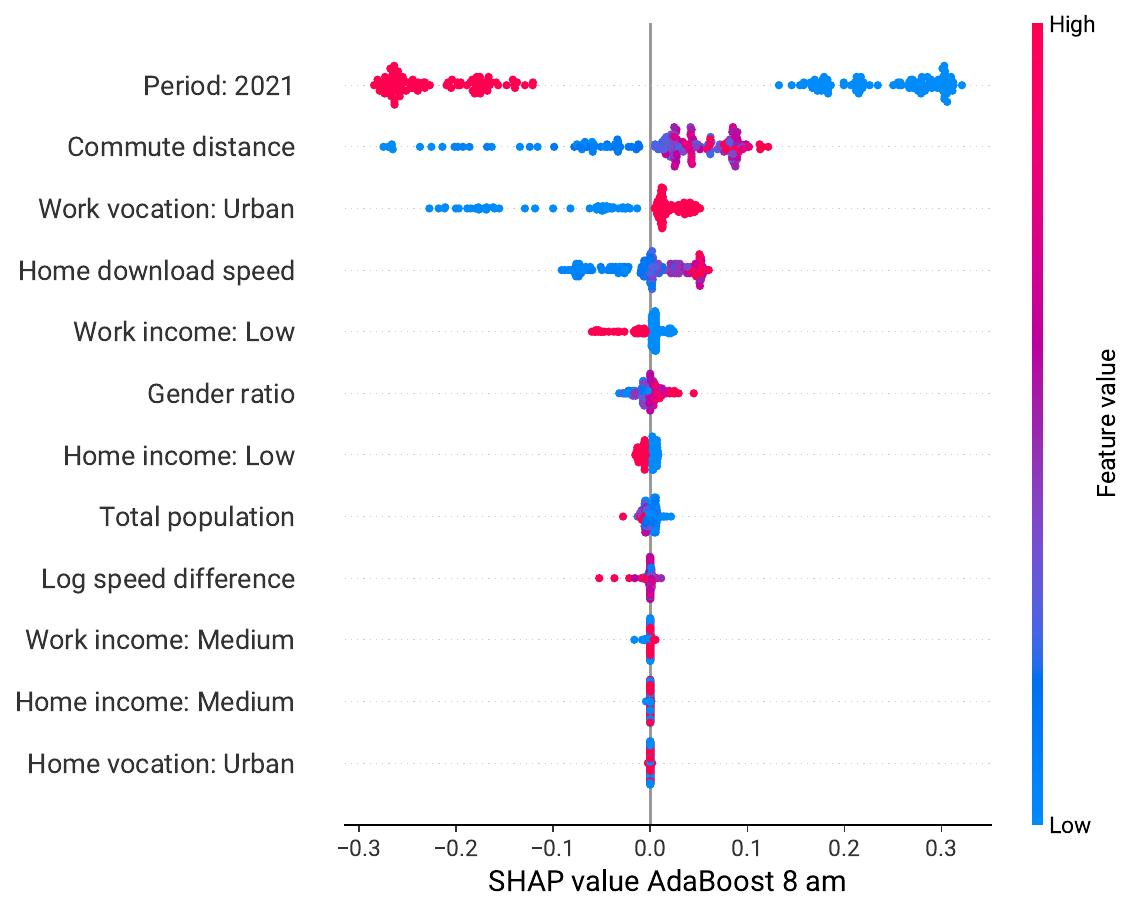}
        \caption{SHAP -- AdaBoost, 08:00}
        \label{fig:shap_ada_8}
    \end{subfigure}
    \hfill
    \begin{subfigure}[b]{0.32\textwidth}
        \centering
        \includegraphics[width=\textwidth]{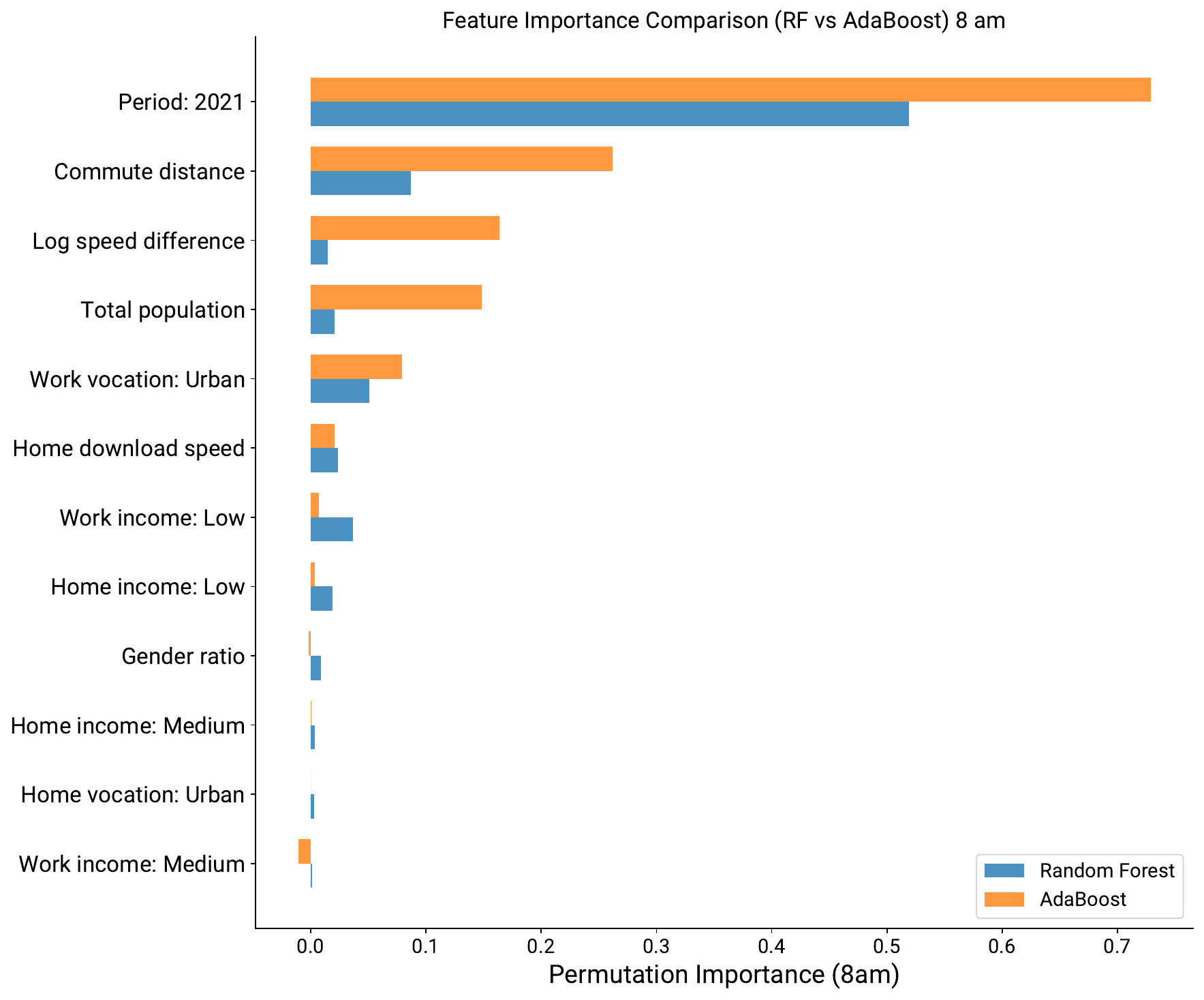}
        \caption{Permutation importance, 08:00}
        \label{fig:imp_8}
    \end{subfigure}

    \vspace{0.5em}

    \begin{subfigure}[b]{0.32\textwidth}
        \centering
        \includegraphics[width=\textwidth]{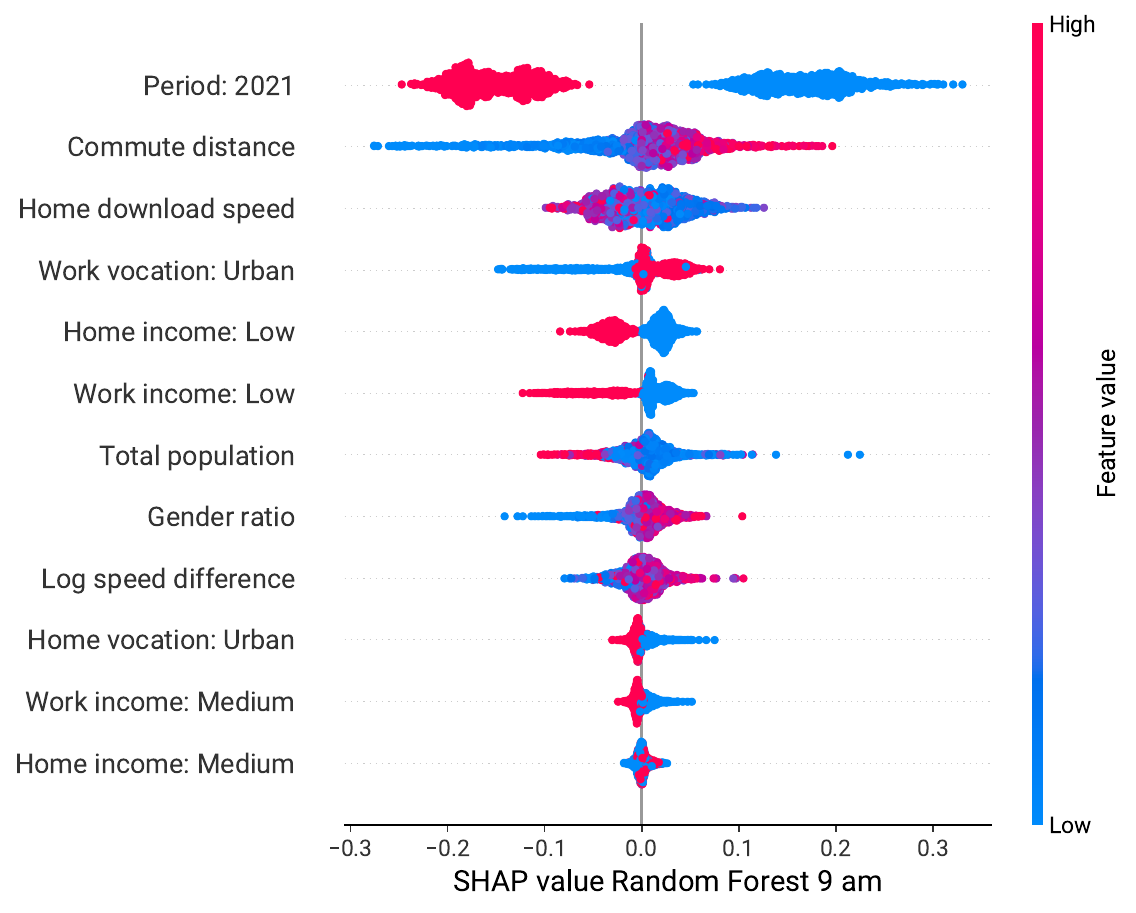}
        \caption{SHAP -- RF, 09:00}
        \label{fig:shap_rf_9}
    \end{subfigure}
    \hfill
    \begin{subfigure}[b]{0.32\textwidth}
        \centering
        \includegraphics[width=\textwidth]{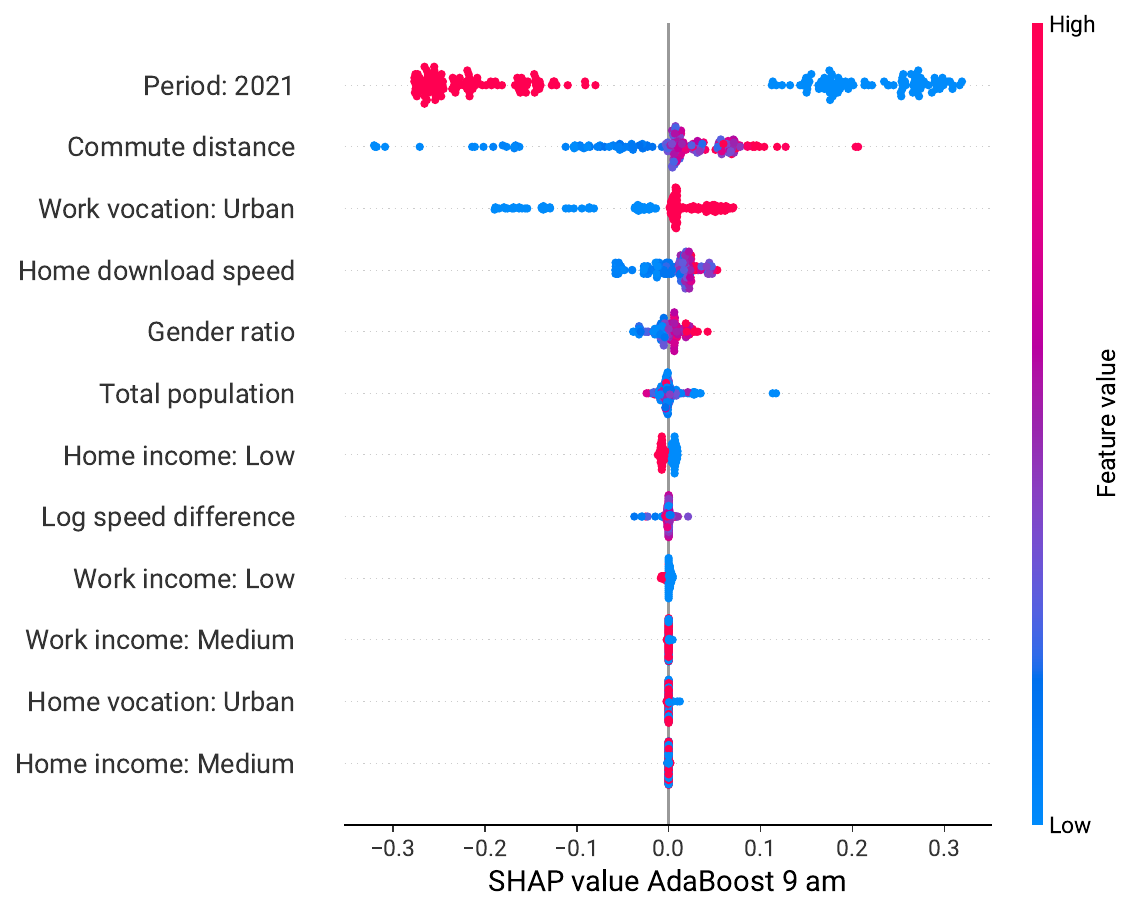}
        \caption{SHAP -- AdaBoost, 09:00}
        \label{fig:shap_ada_9}
    \end{subfigure}
    \hfill
    \begin{subfigure}[b]{0.32\textwidth}
        \centering
        \includegraphics[width=\textwidth]{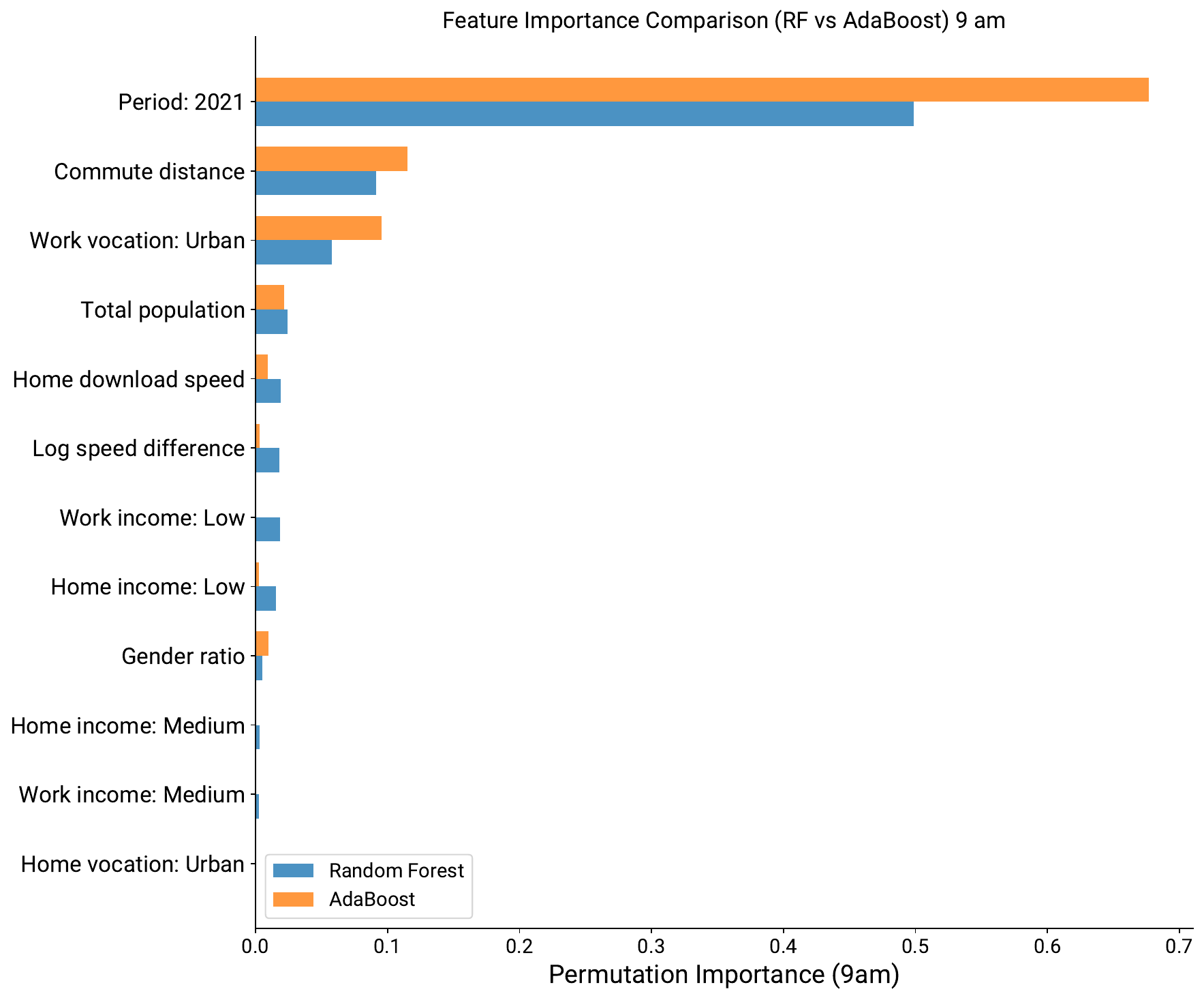}
        \caption{Permutation importance, 09:00}
        \label{fig:imp_9}
    \end{subfigure}

    \caption{\textbf{Feature importance across working hours ndefinitions.} Each row corresponds to a working hour window definition (07:00, 08:00, 09:00). Left column: SHAP summary plots for Random Forest, showing the distribution of SHAP values, with color encoding reflecting feature value magnitude. Center column: SHAP summary plots for AdaBoost. Right column: permutation importance on the held-out test set for both models (Random Forest in blue, AdaBoost in orange), averaged over 30 repeated permutations.}
    \label{fig:shap_importance_all}
\end{figure}

As a complementary measure of feature relevance, we computed permutation importance for both models on the held-out test set~\cite{shap_permutation_importance_docs}. Permutation importance quantifies the drop in predictive accuracy when a feature's values are randomly shuffled, thereby breaking its association with the outcome, averaged over repeated permutations, to reduce variance. Unlike SHAP values, permutation importance reflects marginal predictive contributions on unseen data rather than in-sample decompositions, making it a useful cross-check. Results from both methods are largely consistent across working hours definitions (see Tab.~\ref{tab:perm_imp} and Fig.~\ref{fig:shap_importance_all}) with the period indicator and commuting distance ranking highest in all cases, and income and speed variables contributing though less strongly.

While the features' importance analysis is broadly consistent with the regression results, we also highlight some differences in relative magnitude. In line with the regression shown in the main, the period indicator confirms its central role in explaining commuting reduction, emerging as the most important predictor overall. The second most important feature is commuting distance. In the multivariate regression analysis, commuting distance is significantly positive, but its coefficient is not among the largest in absolute value. In contrast, the SHAP analysis assigns it a higher relative importance, suggesting that its contribution to predictive performance is stronger than what is implied by the linear coefficient alone.

By comparison, income-related variables have lower relevance in these models based on decision trees. While some income categories were significant in the regression, their feature importance is comparatively lower. Nevertheless, the different models are largely consistent regarding primary effects, although they differ in how the absolute importance of secondary predictors.

\begin{table}[ht]
\centering
\caption{Mean permutation importance on the test set for Random Forest (RF) and AdaBoost (Ada) across working hours definitions. Features are sorted by average importance across the three windows.}
\label{tab:perm_imp}
\begin{tabular}{lcccccc}
\hline
 & \multicolumn{2}{c}{09:00} & \multicolumn{2}{c}{08:00} & \multicolumn{2}{c}{07:00} \\
Feature & RF & Ada & RF & Ada & RF & Ada \\
\hline
Period: 2021            & 0.498 & 0.677 & 0.519 & 0.729 & 0.493 & 0.689 \\
Commute distance        & 0.091 & 0.115 & 0.087 & 0.262 & 0.099 & 0.117 \\
Work vocation: Urban    & 0.058 & 0.095 & 0.051 & 0.079 & 0.063 & 0.094 \\
Home download speed     & 0.019 & 0.009 & 0.024 & 0.021 & 0.034 & 0.009 \\
Total population        & 0.024 & 0.021 & 0.021 & 0.149 & 0.007 & $-0.012$ \\
Work income: Low        & 0.019 & 0.000 & 0.036 & 0.007 & 0.018 & 0.004 \\
Log speed difference    & 0.018 & 0.003 & 0.015 & 0.164 & 0.006 & $-0.006$ \\
Home income: Low        & 0.015 & 0.002 & 0.019 & 0.004 & 0.014 & 0.005 \\
Gender ratio            & 0.005 & 0.010 & 0.009 & $-0.002$ & $-0.002$ & $-0.009$ \\
Home income: Medium     & 0.003 & 0.001 & 0.004 & 0.000 & 0.001 & 0.000 \\
Work income: Medium     & 0.003 & 0.000 & 0.001 & $-0.011$ & 0.002 & 0.001 \\
Home vocation: Urban    & 0.001 & 0.000 & 0.003 & 0.000 & 0.004 & 0.000 \\
\hline
\end{tabular}
\end{table}

\end{document}